\documentclass[prd,aps,twocolumn,showpacs,nofootinbib]{revtex4-1}

\usepackage{amsfonts,color,graphicx,epsfig,amsmath,multirow}

\usepackage[colorlinks=true,linkcolor=blue,citecolor=blue, urlcolor=blue]{hyperref}

\newcommand{\bfk}{{\bf k}_{\bot}}
\newcommand{\e}{\e^{*(\lambda)}}
\newcommand{\rhob}{^\bot_{2;\rho}}
\newcommand{\wt}{\widetilde}
\newcommand{\nn}{\nonumber}
\newcommand{\Vub}{|V_{\rm ub}|}

\begin{document}

\title{$B\to \rho$ transition form factors and the $\rho$-meson transverse leading-twist distribution amplitude}

\author{Hai-Bing Fu}
\author{Xing-Gang Wu}
\email{Email: wuxg@cqu.edu.cn}
\author{Hua-Yong Han}
\author{Yang Ma}
\address{Department of Physics, Chongqing University, Chongqing 401331, P.R. China}
\address{Institute of Theoretical Physics, Chongqing University, Chongqing 401331, P.R. China}

\date{\today}

\begin{abstract}
The QCD light-cone sum rules (LCSR) provides an effective way for dealing with heavy-to-light transition form factors (TFFs), in which the non-perturbative dynamics are parameterized into the light-meson's light-cone distribution amplitudes (LCDAs). By taking the chiral correlator as the starting point, we derive new LCSRs for $B\to\rho$ TFFs up to twist-4 accuracy. In those LCSRs, the twist-2 transverse LCDA $\phi_{2;\rho}^\bot$ provides dominant contribution, while the twist-3 and twist-4 contributions are $\delta^2$-suppressed ($\delta \simeq m_\rho/m_b$). Thus, they provide good platforms for testing $\phi_{2;\rho}^\bot$ behavior. For the purpose, we suggest a WH-model for $\phi_{2;\rho}^\bot$, in which a single parameter $B_{2;\rho}^\bot\sim a^\bot_2$ dominantly controls its longitudinal distribution. When setting $B_{2;\rho}^\bot\in[-0.20,+0.20]$, $\phi_{2;\rho}^\bot$ varies from the single-peak behavior to the double-humped behavior. We present a detailed comparison of the LCSR estimation for $B\to\rho$ TFFs with those predicted by pQCD and Lattice QCD calculations. The TFFs become smaller with the increment of $B_{2;\rho}^\bot$ and a larger $B_{2;\rho}^\bot$ is not allowed by lattice QCD predictions. By using the extrapolated TFFs, we further predict the CKM-matrix element $|V_{\rm ub}|$ with the help of two $B\to\rho$ semi-leptonic decays. The predicted $|V_{\rm ub}|$ increases with the increment of $B_{2;\rho}^\bot$, i.e. we have $\Vub=(2.91\pm0.19)\times 10^{-3}$ for $B_{2;\rho}^\bot=-0.20$ and $\Vub=(3.11\pm0.19)\times 10^{-3}$ for $B_{2;\rho}^\bot=0.00$. If treating the BABAR prediction on $|V_{\rm ub}|$ as a criteria, we observe that $B^{\perp}_{2;\rho}$ should be within the region of $[-0.20,+0.10]$, which indicates that the $\rho$-meson LCDA prefers a single-peak behavior rather than a double-humped behavior.

\end{abstract}


\maketitle

\section{Introduction}\label{sec:I}

The $B\to \rho$ transition form factors (TFFs) are key components for the semileptonic decay $B\to \rho\ell\bar\nu_\ell$ or the rare penguin-induced flavor-changing-neutral-current decays $B\to \rho\gamma$ and $B\to\rho\ell^+\ell^-$. Since the leptons do not participate in strong interaction, the $B$-meson semileptonic decays provide much clearer samples than its hadronic decays and are important for studying strong and weak interactions. Those decays provide good platform for studying the $\rho$-meson light-cone distribution amplitudes (LCDAs) and the Cabibbo-Kobayashi-Maskawa (CKM) matrix elements.

The QCD light-cone sum rules (LCSR)~\cite{LCSR:1,LCSR:2} provides an efficient tool in making predictions for exclusive processes. This method is based on the operator product expansion (OPE) near the light cone $x^2 \rightsquigarrow 0$. Different to the conventional SVZ sum rules~\cite{svz}, the LCSR expanses the operator over its twists rather than its dimensions. Within the framework of LCSR, all non-perturbative dynamics is parameterized into LCDAs instead of the quark and gluon condensates. Since its invention, the LCSR has been widely adopted for studying $B\to$ light meson decays, cf.Refs.\cite{Brho,Ball05,B:3,B:4,chiral,Wu2009}. It is noted that the LCSRs for $B\to\rho$ TFFs $A_0(q^2)$, $A_1(q^2)$, $A_2(q^2)$ and $V(q^2)$ are applicable for low and intermediate $q^2$ region. Here $q$ stands for the momentum transfer between the $B$-meson and the $\rho$-meson. This can be compared with the applicability of high $q^2$ region for the lattice QCD~\cite{Lattice96:1,Lattice96:2,Lattice98,Lattice04} and low $q^2$ region for the pQCD~\cite{LuCD:pQCD,LiHN:2002}. Those approaches are complementary to each other, and by combining the results derived from those approaches, one may obtain a full understanding of $B\to\rho$ TFFs in whole $q^2$ region. Particularly, the LCSR provides a link between the pQCD and the lattice QCD estimations. Thus, a better LCSR prediction with less uncertainties is helpful for a better understanding of those TFFs. Furthermore, by utilizing the LCSR predictions for $B\to\rho$ TFFs, one may inversely get interesting information on the QCD parameters such as the $\rho$-meson LCDAs via a detailed comparison with the lattice QCD estimation and the experimentally known hadronic parts.

The meson's LCDA, which relates to the matrix elements of the nonlocal light-ray operators sandwiched between the hadronic state and the vacuum, can exhibit all the meson's properties. It also provides underlying links between the hadronic phenomena at the small and large distances. The structures of the $\rho$-meson's LCDAs are much more complex than the pseudoscalar pion or kaon LCDAs. There are chiral-even or chiral-odd LCDAs for the vector $\rho$-meson due to chiral-even and chiral-odd operators in the matrix elements. The $\rho$-meson has two polarization states, either longitudinal ($\|$) or transverse ($\perp$), which may correspond to different twist structures. As suggested by Refs.\cite{Ball04:Brho,P.Ball:1998-1999}, it is convenient to rearrange the $\rho$-meson LCDAs via a parameter $\delta$, i.e. $\delta \simeq m_\rho/m_b$. Following the idea, a classification of the twist-2, twist-3 and twist-4 LCDAs up to $\delta^3$-order are collected in Table \ref{DA_delta}.

\begin{table}[tb]
\centering
\begin{tabular}{ | c | c| c | c | c | }
\hline
 twist & ~~$\delta^0$~~  & ~~$\delta^1$~~ & ~~$\delta^2$~~ & ~~$\delta^3$~~   \\
\hline
~~2~~  & $\phi_{2;\rho}^\bot$ & $\phi_{2;\rho}^\|$ & / & / \\
\hline
~~3~~  & / & $\phi_{3;\rho}^\bot$,$\psi_{3;\rho}^\bot$,$\Phi_{3;\rho}^\|$, $\widetilde\Phi_{3;\rho}^\|$ & $\phi_{3;\rho}^\|, \psi_{3;\rho}^\|, \Phi_{3;\rho}^\bot$ & / \\
\hline
~~4~~  & / & / & $  \phi_{4;\rho}^\bot$,$\psi_{4;\rho}^\bot$, $\Psi_{4;\rho}^\bot$,$\widetilde{\Psi} _{4;\rho}^\bot$ & $\phi_{4;\rho}^\|$,$\psi_{4;\rho}^\|$ \\
\hline
\end{tabular}
\caption{The $\rho$-meson LCDAs with different twist-structures up to $\delta^3$-order~\cite{P.Ball:1998-1999,Ball04:Brho}, where $\delta \simeq m_\rho/m_b$.} \label{DA_delta}
\end{table}

By using standard correlator to deal with $B\to\rho$ TFFs, it has been found that all the twist-2, twist-3 and twist-4 LCDAs listed in Table \ref{DA_delta} are in the resultant LCSRs. Their relative importance follow the $\delta$-counting rule suggested by Refs.\cite{Ball04:Brho,P.Ball:1998-1999}. All $\rho$-meson LCDAs, especially those of $\delta^{1}$-order, have sizable contributions, then they should be taken into consideration for a sound prediction~\cite{Ball1997:Brho,Ball04:Brho}. Thus, the accuracy of the LCSR depends heavily on how well we know those $\delta^0$- and $\delta^1$- LCDAs. However, at the present, all $\rho$-meson LCDAs are far from affirmation, then, it is helpful to find a proper way to suppress those uncertain sources as much as possible so as to achieve a more reliable prediction. In the paper, we shall show that the LCSR derived with the help of a chiral current~\cite{chiral,ChiralCurrent:1,ChiralCurrent:2, ChiralCurrent:3,Wu2009} can be adopted for such purpose.

Under such approach, a proper chiral correlator shall be adopted to derive the LCSR such that the twist-3 LCDAs $\phi_{3;\rho}^\bot$, $\psi_{3;\rho}^\bot$, $\Phi_{3;\rho}^\|$, $\widetilde\Phi_{3;\rho}^\|$ and the parallel twist-2 LCDA $\phi_{2;\rho}^\|$ make no contributions to $B\to\rho$ TFFs. The leaving twist-3 and twist-4 LCDAs are $\delta^2$-suppressed, thus, the errors due to those uncertain high-twist LCDAs themselves shall be largely suppressed. The $\rho$-meson transverse leading-twist LCDA $\phi_{2;\rho}^\bot$ is at $\delta^{0}$-order and provides the dominant contribution to $B\to\rho$ TFFs. This, inversely, makes the $B\to \rho$ semi-leptonic decays be good places for testing different models of $\phi_{2;\rho}^\bot$. For the purpose, we shall first construct a general model for $\rho$-meson leading-twist LC wavefunction (LCWF) $\psi_{2;\rho}^\bot$ based on the well-known Brodsky-Huang-Lepage (BHL) prescription~\cite{BHL}. As will be shown latter, its longitudinal behavior is dominantly controlled by an input parameter $B\rhob$, i.e. we have $B\rhob\sim a^{\bot}_2$ with $a^{\bot}_2$ being the second Gegenbauer moment of $\phi_{2;\rho}^\bot$. In the literature, several LCDA models have been suggested, either the one in Gegenbauer expansion~\cite{Ball04:Brho,CZlike:Report} or the one based on AdS/QCD theory~\cite{Brodsky:2007hb,Brodsky:2008pg}. Then, it is helpful to make a comparison of all those models and show how they affect the LCSRs for $B\to\rho$ TFFs.

The LCSR can be extrapolated to any physical region, then we can compare our LCSR prediction for $B\to\rho$ TFFs with the lattice QCD predictions~\cite{Lattice96:1,Lattice96:2,Lattice98,Lattice04}. Through such a comparison, one may determine a possible range for $B\rhob$ and a possible behavior for $\phi_{2;\rho}^\bot$. Furthermore, the CKM matrix element $|V_{\rm ub}|$ has been studied by various experimental groups by measuring $B\to\rho$ semi-leptonic decays~\cite{BABAR:2010,BABAR:2005,BABAR:2003,CLEO:1999}. It is also helpful to compare the LCSR predictions on the CKM matrix element $|V_{\rm ub}|$ with the experimental predictions.

The remaining parts of the paper are organized as follows. In Sec.\ref{sec:II}, we present the formulas for $B\to\rho$ semi-leptonic decay, and we present the calculation technology for $B\to\rho$ TFFs under the LCSR approach. In Sec.\ref{sec:III}, we present our numerical results and discussions. First, we study the properties of several models for $\rho$-meson leading-twist LCDA $\phi_{2;\rho}^\bot$. Second, we study the $B\to\rho$ TFFs under various LCDA models and make a comparison of those TFFs with the lattice QCD estimation. Finally, we present a comparison of the LCSR prediction for $\Vub$ with the experimental estimations. Sec.\ref{sec:Summary} is reserved for a summary.

\section{Calculation Technology} \label{sec:II}

The key component of the semileptonic decay $B\to\rho\ell\nu_\ell$ is the matrix element $\langle \rho (p,\lambda )|\bar q{\gamma _\mu }(1 - {\gamma _5})b |B(p+q)\rangle$, which can be expanded as
\begin{eqnarray}
&&\langle \rho (p,\lambda )|\bar q{\gamma _\mu }(1 - {\gamma _5})b|B(p+q)\rangle \nn\\
&=& - ie_\mu ^{*(\lambda )}(m_B + m_\rho )A_1(q^2) \nn\\
&&+ i(e^{*(\lambda )}\cdot q)\frac{A_2(q^2)(2p + q)_\mu}{m_B + m_\rho }\nn\\
&&+ iq_\mu (e^{*(\lambda )}\cdot q)\frac{2 m_\rho } {q^2}[A_3(q^2) - A_0(q^2)]\nn \\
&&+\epsilon_{\mu\nu\alpha\beta}e^{*(\lambda )\nu} q^\alpha p^\beta \frac{2V(q^2)}{m_B + m_\rho }, \label{Brho:matrix}
\end{eqnarray}
where $e^{(\lambda)}$ stands for $\rho$-meson polarization vector with $\lambda$ being its transverse ($\bot$) or longitudinal ($\|$) component, respectively. $p$ is the $\rho$-meson momentum and $q = p_B - p_\rho$ is the four-momentum transfer between those two mesons. $A_{i}$ with $i=(0,\cdots,3)$ and $V$ are $B\to\rho$ TFFs, in which $A_{1,2,3}$ satisfy the following relation
\begin{equation}
A_3(q^2)=\frac{m_B+m_\rho}{2m_\rho}A_1(q^2)-\frac{m_B-m_\rho}{2m_\rho}A_2(q^2). \label{A3}
\end{equation}
At the endpoint, we have $A_0(0)=A_3(0)$.

If the leptonic mass can be ignored ($\ell=e$ or $\mu$), due to chiral suppression, the total differential decay width for $B\to\rho\ell\nu_\ell$ can be written as
\begin{equation}
\frac{d\Gamma}{dq^2}={\cal G}\Vub^2 \lambda(q^2)^{1/2} q^2[H_0^2(q^2)+H_+^2(q^2)+H_-^2(q^2)], \label{difftot}
\end{equation}
where ${\cal G}={G_F^2}/{(192\pi^3 m_B^3)}$ with the fermi coupling constant $G_F=1.166\times10^{-5}\;{\rm GeV}^{-2}$, and the phase-space factor $\lambda(q^2)=(m_B^2 + m_\rho^2 - q^2)^2-4 m_B^2 m_\rho^2$. Here, the transverse and longitudinal helicity amplitudes $H_{0,\pm}(q^2)$ are given by
\begin{widetext}
\begin{eqnarray}
H_\pm(q^2) &=& (m_B+m_\rho)A_1(q^2)\mp\frac{\lambda(q^2)^{1/2}} {m_B+m_\rho}V(q^2), \label{Hpm} \\
H_0(q^2) &=& \frac{1}{2m_\rho(q^2)^{1/2}}\bigg\{(m_B^2-m_\rho^2-q^2) (m_B+m_\rho) \times A_1(q^2)-\frac{\lambda(q^2)}{m_B+m_\rho}A_2(q^2)\bigg\}. \label{H0}
\end{eqnarray}
\end{widetext}
In the helicity basis, each TFF corresponds to a transition amplitude with definite spin-parity quantum numbers in the lepton pair center-of-mass frame. This relates the TFFs $A_1$, $A_2$ and $V$ to the total angular momentum and parity quantum numbers of the $B-\rho$ meson pair, i.e. $J^P = 1^+$, $1^+$ and $1^-$~\cite{Ball1997:Brho}, respectively. The physical region for the squared four-momentum transfer is $0\leq q^2 \leq q_{\rm max}^2\equiv(m_B-m_\rho)^2$.

In most of the kinematic region, the $B\to\rho$ TFFs are non-perturbative, which can only be treated within some non-perturbative approaches such as the QCD LCSR and the lattice QCD approach. In low and intermediate energy region, it can be calculated within the framework of LCSR. How to ``design" a proper correlator for a particular case is the key but non-trivial problem for LCSR. By a suitable choice of the correlator, one can not only obtain the right properties of the hadrons/TFFs but also greatly simplify the theoretical uncertainties.

For the $B\to\rho$ TFFs, one needs to deal with the following correlator:
\begin{widetext}
\begin{eqnarray}
\Pi _\mu (p,q) &=&  i\int d^4x e^{iq\cdot x}\langle\rho (p,\lambda)|{\rm T} \left\{\bar q_1(x)\gamma_\mu(1-\gamma_5) b(x), j_B^\dag (0)\right\} |0\rangle \label{correlator:1}\\
&=& \Pi_1 e_\mu^{*(\lambda )}-\Pi_2(e^{*(\lambda )}\cdot q)(2p + q)_\mu - \Pi_3(e^{*(\lambda )}\cdot q)q_\mu + i\Pi _V\epsilon_\mu^{\alpha\beta\gamma}e_\alpha ^{*(\lambda )}q_\beta p_\gamma . \label{correlator:2}
\end{eqnarray}
\end{widetext}
The current $j_B^\dag (x)$ can be conventionally and simply chosen as $i \bar b(x) \gamma_5 q_2(x)$ such that it has the same quantum state as that of the pseudoscalar $B$-meson with $J^{P}=0^-$. As mentioned in the Introduction, such a choice of correlator shall result in a complex series with all the possible $\rho$-meson twist-structures~\cite{Ball1997:Brho,Ball04:Brho}. Large theoretical uncertainties shall be introduced due to unknown/less-known LCDAs. On the other hand, it is convenient to choose $j_B^\dag (x)$ as a chiral current, either $i \bar b(x)(1- \gamma_5)q_2(x)$ or $i \bar b(x)(1+ \gamma_5)q_2(x)$, to do the calculation. The advantage of such a choice lies in that one can highlight the contributions from different twists of $\rho$-meson DAs to the TFFs by selecting a proper chiral current. Then the properties of those highlighted LCDAs can be tested with a much higher precision. More explicitly, by taking $j_B^\dag (x)= i \bar b(x)(1- \gamma_5)q_2(x)$, we can highlight the chiral-even $\rho$-meson DAs' contributions; while by taking $j_B^\dag (x)= i \bar b(x)(1 + \gamma_5)q_2(x)$, one can highlight the chiral-odd $\rho$-meson DAs' contributions.

In the present paper, we shall adopt $j_B^\dag (x)= i \bar b(x)(1 + \gamma_5)q_2(x)$ to deal with the $B\to\rho$ TFFs \footnote{The cases for $j_B^\dag (x)= i \bar b(x)(1- \gamma_5)q_2(x)$ can be found in Ref.\cite{fu1}.}. Under such choice, it is noted that the resulting hadronic representation of the correlator depends not only on the resonances with $J^P=0^-$ state but also on those of $J^P = 0^+$ state. This is the price of introducing a chiral correlator for LCSR. But it is worthwhile, since we can eliminate the large uncertainties from the twist-2 and twist-3 LCDAs at $\delta^1$-order, and we may avoid the pollution from the scalar resonances with $J^P = 0^+$ by choosing proper continuum threshold $s_0$. Our final results with slight $s_0$ dependence also confirm this assumption.

The correlator, Eq.(\ref{correlator:1}), is an analytic function of $q^2$ defined at both negative (space-like) and positive (time-like) values of $q^2$. In the time-like region, the long-distance quark-gluon interactions become important and, eventually, the quarks form hadrons. To deal with the correlator in the time-like region, one can insert a complete series of intermediate hadronic states with the same quantum numbers as the current operator $\bar bi(1+\gamma_5)q_2$ to obtain the hadronic representation. After isolating the pole term of the lowest pseudoscalar $B$-meson, we obtain
\begin{widetext}
\begin{eqnarray}
\Pi_\mu^{\rm H}(p,q) =\frac{\langle\rho(p,\lambda)|\bar q_1 \gamma _\mu (1 - \gamma_5)b|B\rangle \langle B|\bar b(0)i \gamma_5 q_2(0)|0\rangle }{m_B^2 - (p + q)^2} + \sum\limits_{\rm H} \frac{\langle \rho(p,\lambda)|\bar q_1\gamma_\mu (1 - \gamma_5)b|B^{\rm H}\rangle\langle B^{\rm H}|\bar b i (1 + \gamma _5)q_2|0\rangle }{m_B^{{\rm H}2} - (p + q)^2},  \label{Hadronic_expression}
\end{eqnarray}
\end{widetext}
where $\langle B|\bar b i \gamma_5 q_2|0\rangle=m_B^2 f_B/m_b$ with $f_B$ standing for the $B$-meson decay constant. Thus, the hadronic expressions for the correlator are
\begin{eqnarray}
\Pi_1^{\rm H}[q^2,(p + q)^2] &&= \frac{m_B^2f_B(m_B + m_\rho )}{m_b[m_B^2 -(p + q)^2]}A_1(q^2) \nn\\
&& + \int_{s_0}^\infty  \frac{\rho_1^{\rm H}}{s - (p + q)^2} ds, \label{Hadronic:A1}
\end{eqnarray}
\begin{eqnarray}
\Pi _2^{\rm H}[q^2,(p + q)^2] &&= \frac{m_B^2 f_B}{m_b(m_B + m_\rho )[m_B^2 - (p + q)^2]}\nn\\
&& \times A_2(q^2)+ \int_{s_0}^\infty  \frac{\rho_2^{\rm H}} {s - (p + q)^2} ds,\label{Hadronic:A2}
\end{eqnarray}
\begin{eqnarray}
\Pi_V^{\rm H}[q^2,(p + q)^2] &&= \frac{2 m_B^2 f_B}{m_b(m_B + m_\rho )[m_B^2 - (p + q)^2]}\nn\\
&& \times V(q^2) + \int_{s_0}^\infty  \frac{\rho_V^{\rm H}}{s - (p + q)^2} ds.\label{Hadronic:V}
\end{eqnarray}
The contributions from higher resonances and the continuum states above $s_0$ have been written in terms of dispersion integrations. The spectral densities $\rho^H_i(s)$ can be approximated via the quark-hadron duality ansatz~\cite{svz}, $\rho^H_i(s) = \rho^{\rm QCD}_i(s)\theta(s-s_0)$. As shown by Eqs.(\ref{Hpm},\ref{H0}), the decay width for $B\to\rho l\nu$ with massless leptons involves $A_1$, $A_2$ and $V$ only, so we present the procedures on how to derive the LCSRs of those TFFs in detail. The LCSR for the fourth independent TFF $A_0$ can be achieved via a similar way.

On the other hand, within the space-like region, we can calculate the correlator via the QCD theory. In large space-like region, which corresponds to small light-cone distance $x^2\rightsquigarrow0$, we have $(p+q)^2-m_b^2\ll 0$ with the momentum transfer $q^2 \sim {\cal O}(1\;{\rm GeV}^2)\ll m^2_b$. In this region, the correlator can be treated by the operator product expansion (OPE) with the coefficients being pQCD calculable. As a basis, we adopt the following $b$-quark propagator to do the calculation
\begin{widetext}
\begin{displaymath}
\langle 0|{\rm T} \{ b(x)\bar b(0)\} |0\rangle  = i\int \frac{d^4 k} {(2\pi )^4} e^{-ik\cdot x} \frac{\not\! k + m_b}{m_b^2 - k^2}- i g_s \int \frac{d^4 k}{(2\pi )^4} e^{- ik\cdot x} \int_0^1 dv G^{\mu \nu }(vx) \left[ \frac{1}{2}\frac{\not\! k + m_b} {(m_b^2 - k^2)^2} \sigma_{\mu\nu} + \frac{v}{m_b^2 - k^2} x_\mu \gamma_\nu \right], \label{sq1}
\end{displaymath}
\end{widetext}
where $G_{\mu\nu}$ is the gluonic field strength and $g_s$ denotes the strong coupling constant. After applying the OPE to the correlator, we obtain
\begin{widetext}
\begin{eqnarray}
\Pi_\mu^{\rm OPE}&& = \int \frac{d^4x d^4k}{(2\pi)^4} e^{i(q - k)\cdot x} \bigg\{ \frac{1}{m_b^2 - k^2}\bigg\{ 2k^\mu \langle \rho(p,\lambda)|\bar q_1(x) q_2(0)|0\rangle  - 2i{k^\nu }\langle \rho (p,\lambda)|\bar q_1(x)\sigma _{\mu \nu} q_2(0)|0\rangle \nn\\
&& - \epsilon_{\mu\nu\alpha\beta} k^\nu \langle \rho (p,\lambda)|\bar q_1(x)\sigma_{\alpha\beta} q_2(0)|0\rangle \bigg\} - \int dv \bigg\{ \frac{k^\nu}{(m_b^2 - k^2)^2} \bigg[ - i\langle\rho(p,\lambda)|\bar q_1(x) g_s G_{\mu\nu}(vx)q_2(0)|0\rangle \nn\\
&&- 2\langle\rho(p,\lambda)|\bar q_1(x)\sigma_{\mu\alpha} g_s G^{\alpha \nu}(vx) q_2(0)|0\rangle  + 2i\langle\rho(p,\lambda)| \bar q_1(x)i g_s {\wt G}_{\mu\nu}(vx) \gamma_5 q_2(0)|0\rangle \bigg] \nn\\
&& + \frac{2v x_\alpha }{m_b^2 - k^2} \left[-\langle\rho(p,\lambda)|\bar q_1(x)g_s G_{\mu\alpha}(vx) q_2(0)|0\rangle - i\langle\rho(p,\lambda)|\bar q_1(x)\sigma_{\mu\beta} g_s G^{\alpha\beta}(vx) q_2(0)|0\rangle \right]\bigg\}\bigg\},\label{correlator2}
\end{eqnarray}
\end{widetext}
where $\wt G_{\mu \nu}(vx) = \epsilon_{\mu \nu \alpha \beta} G^{\alpha \beta }(vx)/2$. We need to deal with the meson-to-vacuum transition matrix element with different $\gamma$-structures, $\Gamma={\bf 1}$, $i\gamma_5$ and $\sigma_{\mu\nu}$, for two- and three-particle LCDAs' contributions, respectively. It is noted that the meson-to-vacuum matrix element with $\Gamma=i\gamma_5$ vanishes for two-particle contribution, because it is impossible to construct a pseudoscalar quantity from $p_\mu$, $x_\mu$ and $e_\mu^{\lambda}$. Up to twist-4 accuracy, those transition matrix elements can be expanded as~\cite{P.Ball:1998-1999}:
\begin{widetext}
\begin{eqnarray}
\langle \rho (p,\lambda)|\bar q_1(x)\sigma_{\mu \nu}q_2(0)|0\rangle &=& - i f_\rho^\bot \int_0^1 du e^{iu(p\cdot x)}\bigg\{(e_\mu^{*(\lambda )}p_\nu - e_\nu^{*(\lambda )}p_\mu)\bigg[\phi_{2;\rho}^\bot(u)+ \frac{m_\rho^2 x^2}{16} \phi_{4;\rho}^\bot(u)\bigg]  \nn\\
&& +(p_\mu x_\nu - p_\nu x_\mu ) \frac{e^{*(\lambda)}\cdot x}{(p\cdot x)^2} m_\rho^2 \left[\phi_{3;\rho}^\|(u) - \frac{1}{2}\phi_{2;\rho}^\bot(u) - \frac{1}{2} \psi_{4;\rho}^\bot(u)\right] \nn\\
&& +\frac{1}{2}\left(e_\mu^{*(\lambda )}{x_\nu } - e_\nu^{*(\lambda)}x_\mu\right) \frac{m_\rho^2}{p\cdot x} \left[ \psi_{4;\rho}^\bot(u)-\phi_{2;\rho}^\bot(u) \right] \bigg\},\label{DA1} \\
\langle \rho (p,\lambda )|\bar q_1(x) q_2(0)|0\rangle &=& - \frac{i}{2} f_\rho^\bot\left(e^{*(\lambda)}\cdot x\right) m_\rho^2\int_0^1 du e^{iu(p\cdot x)} \psi_{3;\rho }^\parallel(u),  \label{DA2} \\
\langle\rho(p,\lambda)|\bar q_1(x)\sigma_{\alpha\beta}g_s G^{\mu\nu}(vx)q_2(0)|0\rangle &=& m_\rho^2 f_\rho^\bot \frac{e^{*(\lambda )}\cdot x}{2(p \cdot x)}\left[p_\mu\left(p_\alpha g_{\beta\nu}^\bot - {p_\beta}g_{\alpha\nu}^\bot\right) - p_\nu\left(p_\alpha g_{\beta\mu}^\bot-p_\beta g_{\alpha \mu }^\bot\right)\right] \Phi_{3;\rho }^\bot(v,p\cdot x), \label{DA4}\\
\langle\rho(p,\lambda)|\bar q_1(x) g_s G^{\mu\nu}(vx)q_2(0)|0\rangle &=& -i m_\rho^2 f_\rho^\bot \left[e_{\bot\mu}^{*(\lambda)}p_\nu - e_{\bot\nu}^{*(\lambda)}p_\mu\right] \Psi_{4;\rho}^\bot(v,p\cdot x), \label{DA5}\\
\langle\rho(p,\lambda)|\bar q_1(x) i{g_s}\tilde G_{\mu\nu} (vx)\gamma_5q_2(0)|0\rangle &=& i m_\rho^2 f_\rho^\bot \left[e_{ \bot \mu}^{*(\lambda )} {p_\nu } - e_{ \bot \nu }^{*(\lambda )}{p_\mu }\right] \tilde \Psi_{4;\rho }^\bot(v,p\cdot x), \label{DA3}
\end{eqnarray}
\end{widetext}
where $f_\rho^\bot$ represents the $\rho$-meson tensor decay constant,
\begin{displaymath}
\langle \rho(p,\lambda)|\bar q_1(0)\sigma_{\mu \nu }q_2(0)|0\rangle =  i f_\rho^\bot ( e_\nu^{(\lambda )}p_\mu-e_\mu^{(\lambda )}p_\nu ),
\end{displaymath}
and we have set
\begin{eqnarray}
g_{\mu\nu}^\bot &=& g_{\mu\nu}-\frac{p_\mu x_\nu + p_\nu x_\mu}{p\cdot x},\nn
\end{eqnarray}
\begin{eqnarray}
&& e_\mu^\lambda = \frac{e^\lambda\cdot x}{p\cdot x} \left( p_\mu - \frac{m_\rho^2}{2(p\cdot x)} x_\mu\right)+ e_{\bot\mu}^\lambda,\nn\\
&& K(v,{p\cdot x}) = \int{\cal D}\alpha e^{ipx(\alpha_1+v\alpha_3)}K(\underline{\alpha}).\nn
\end{eqnarray}
Here ${\cal D}\alpha=d\alpha_1 d\alpha_2 d\alpha_3\delta(1-\alpha_1 -\alpha_2 -\alpha_3)$ and $K(\underline{\alpha})$ stands for the twist-3 or twist-4 DA $\Phi_{3;\rho}^\bot(\underline{\alpha})$, $\Psi_{4;\rho}^\bot(\underline{\alpha})$ or $\wt\Psi_{4;\rho}^\bot(\underline{\alpha})$, respectively, in which $\underline{\alpha}=\{\alpha_1,\alpha_2,\alpha_3\}$ corresponds to the momentum fractions carried by the antiquark, quark and gluon, respectively. As a tricky point, the integration over $x$ within the above equations can be done by transforming $x_\mu$ in the nominator to $i\partial/\partial(up)_\mu$ or equivalently to $-i\partial/\partial q_\mu$, and by transforming $\phi(u)/(p\cdot x)$ to $-i\int_0^u dv\phi(v)\equiv-i\Phi(u)$.

Equaling the correlator within different regions and applying the conventional Borel transformation to suppress the contributions from the unknown continuum states, we obtain the LCSRs for $B\to\rho$ TFFs:
\begin{widetext}
\begin{eqnarray}
f_B A_1 &&(q^2)e^{-m_B^2/M^2} = \frac{m_b m_\rho^2 f_\rho^\bot}{m_B^2(m_B+m_\rho)} \bigg\{ \int_0^1\frac{du}{u}e^{- s(u)/M^2} \bigg\{\frac{\cal C}{u m_\rho ^2}\Theta(c(u,s_0))\phi_{2;\rho}^\bot(u,\mu) +\Theta(c(u,s_0))\psi_{3;\rho}^\|(u) -\frac{1}{4}  \nn\\
&& \times\bigg[ \frac{m_b^2{\cal C}}{u^3M^4} \wt{\wt\Theta}(c(u,s_0)) + \frac{{\cal C}-2m_b^2}{u^2M^2}\wt\Theta(c(u,s_0)) - \frac{1}{u}\Theta(c(u,s_0))\bigg]\phi_{4;\rho}^\bot(u) - 2\bigg[\frac{\cal C}{u^2M^2}\wt\Theta(c(u,s_0))-\frac{1}{u}\Theta(c(u,s_0))\bigg] \nn\\
&&\times I_L(u)-\bigg[\frac{2m_b^2}{uM^2}\wt \Theta(c(u,s_0)) + \Theta(c(u,s_0)) \bigg]H_3(u)\bigg\} + \int{\cal D}\alpha_i\int_0^1{dv}e^{- s(X)/M^2} \Theta(c(X,s_0)) \bigg[\frac{\underline{\cal C}}{2X^3M^2}-\frac{1}{2X^2}\bigg]\nn\\
&&\times\left[(4v-1)\Psi_{4;\rho}^\bot(\underline\alpha) -\wt\Psi_{4;\rho}^\bot(\underline \alpha)\right]\bigg\},
\label{TFF_A1}
\end{eqnarray}
\begin{eqnarray}
f_B A_2&&(q^2)e^{-m_B^2/M^2} = \frac{ m_b(m_B+m_\rho)m_\rho^2 f_\rho^\bot}{m_B^2}\bigg\{\int_0^1 \frac{du}{u}e^{-s(u)/M^2}\bigg\{ \frac{1}{m_\rho^2}\Theta(c(u,s_0)) \phi_{2;\rho}^\bot (u,\mu) - \frac{1}{M^2}\wt\Theta(c(u,s_0))\psi_{3;\rho}^\|(u) \nn\\
&&-\frac{1}{4}\bigg[ \frac{m_b^2}{u^2M^4}\wt{\wt\Theta}(c(u,s_0)) +\frac{1}{uM^2}\wt\Theta(c(u,s_0))\bigg] \phi_{4;\rho}^\bot(u) + 2\bigg[\frac{{\cal C} - 2m_b^2}{u^2M^4}\wt{\wt\Theta}(c(u,s_0))- \frac{1}{uM^2}\wt\Theta(c(u,s_0))\bigg]I_L(u)  \nn\\
&&-\frac{1}{M^2}\wt\Theta(c(u,s_0)) H_3(u)\bigg\} + \int {\cal D}{\alpha_i} \int_0^1 {dv} e^{-{ s}(X)/M^2}\frac{1}{2X^2M^2}\Theta(c(X,s_0)) \Big[(4v-1)\Psi_{4;\rho}^\bot(\underline\alpha)- \wt \Psi_{4;\rho}^\bot(\underline\alpha)\nn\\
&& +4v\Phi_{3;\rho}^\bot(\underline\alpha)\Big] \bigg\},
\label{TFF_A2}  \\
f_B V&&(q^2)e^{-m_B^2/M^2} = \frac{m_b(m_B + m_\rho)f_\rho^\bot}{m_B^2} \int_0^1 \frac{du}{u} e^{- s(u)/M^2} \bigg\{ \Theta(c(u,s_0)) \phi_{2;\rho}^\bot(u,\mu) - \bigg[\frac{m_b^2}{u^2M^4}\wt {\wt\Theta}(c(u,s_0))+\frac{1}{uM^2}\nn\\
&& \times\wt\Theta(c(u,s_0))\bigg]\frac{m_\rho^2}{4}\phi_{4;\rho}^\bot (u)\bigg\}, \label{TFF_V}
\end{eqnarray}
\end{widetext}
where ${\mathcal C} = m_b^2 + {u^2}m_\rho ^2 - {q^2}$, $\underline {\cal C}  = m_b^2 + {X^2}m_\rho ^2 - {q^2}$ and $s(u)=[m_b^2-\bar u(q^2-u m_\rho^2)]/u$, $s(X) = [m_b^2-\bar X(q^2-X m_\rho^2)]/X$, $X=a_1+va_3$, $\bar u=1-u$ and $\bar X = 1-X$. The simplified functions $I_L(u)$ and $H_3(u)$ are defined as
\begin{eqnarray}
I_L(u) &=& \int_0^u dv \int_0^v dw \big[\phi_{3;\rho}^\|(w) -\frac{1}{2} \phi_{2;\rho}^\bot(w) -\frac{1}{2} \psi_{4;\rho}^\bot(w)\big], \nonumber \\
H_3(u) &=& \int_0^u dv \left[\psi_{4;\rho}^{\bot}(v)-\phi_{2;\rho}^\bot(v)\right]. \nonumber
\end{eqnarray}
In deriving the above LCSRs, we need to deal with two typical Borel transformations, i.e. those involving $E/D^m$ and $1/D^n$ ($1\le m,n \le3$) with $D = m_b^2-(up+q)^2$ and $E = u{p^2} + p \cdot q$. We put the needed formulas in the following:
\begin{eqnarray}
&&{\cal \beta}_{M^2} \int_0^1 {du} \frac{f(u)}{D} = \int_0^1 \frac{du}{u} e^{-\frac{s(u)}{M^2}}\mathop\Theta(c(u,s_0))f(u), \nonumber \\
&&{\cal \beta}_{M^2} \int_0^1 {du} \frac{f(u)}{D^2} =\int_0^1 \frac{du}{u^2 M^{2}} e^{-\frac{s(u)}{M^2}}\wt{\Theta}(c(u,s_0))f(u), \nonumber \\
&&{\cal \beta}_{M^2} \int_0^1 {du} \frac{f(u)}{D^3}  = \int_0^1 \frac{du}{2u^3 M^{4}} e^{-\frac{s(u)}{M^2}} \wt{\wt{\Theta}}(c(u,s_0))f(u), \nonumber \\
&&{\cal \beta}_{M^2} \int_0^1 {du} \frac{E}{D}f(u) = \int_0^1 \frac{du}{2u^2}~{\cal C} e^{-\frac{s(u)}{M^2}} \Theta(c(u,s_0))f(u), \nonumber \\
&&{\cal \beta}_{M^2} \int_0^1 {du} \frac{E}{D^2}f(u)  = \int_0^1 \frac{du~{\cal C}}{2u^{3}M^{2}} e^{-\frac{s(u)}{M^2}}\wt\Theta(c(u,s_0)) f(u) \nn\\
&& \quad\quad\quad\quad -\int_0^1 \frac{du}{2 u^2 }e^{- \frac{s(u)}{M^2}} \Theta(c(u,s_0)) f(u), \nonumber \\
&&{\cal \beta}_{M^2} \int_0^1 {du} \frac{E}{D^3}f(u) = \int_0^1 \frac{du~{\cal C}}{4u^{4}M^{4}} e^{-\frac{s(u)}{M^2}} \wt{\wt{\Theta}}(c(u,s_0)) f(u) \nn\\
&& \quad\quad\quad\quad -\int_0^1 \frac{du}{2u^3 M^{2}}e^{ - \frac{s(u)}{M^2}} \wt{\Theta}(c(u,s_0)) f(u) ,
\end{eqnarray}
where ${\cal \beta}_{M^2}= \lim_{q^2,n\to\infty;(q^2/n)=M^2} \frac{(q^2)^{n+1}}{n!} \left(\frac{d}{d q^2}\right)^n$ stands for the Borel transformation, $c(u,s_0)=u s_0 - m_b^2 + \bar u q^2 - u \bar u m_\rho^2$. $\Theta(c(u,s_0))$ is the conventional step function, $\wt\Theta (c(u,s_0))$ and $\wt{\wt\Theta}(c(u,s_0))$ are defined as
\begin{eqnarray}
&&\int_0^1 \frac{du}{u^2 M^2} e^{-s(u)/M^2}\wt\Theta(c(u,s_0))f(u)
\nn\\
&&~~~~= \int_{u_0}^1\frac{du}{u^2 M^2} e^{-s(u)/M^2}f(u) + \delta(c(u_0,s_0)),
\label{Theta1}\\
&&\int_0^1 \frac{du}{2u^3 M^4} e^{-s(u)/M^2}\wt{\wt\Theta}(c(u,s_0))f(u)
\nn\\
&&~~~~= \int_{u_0}^1 \frac{du}{2u^3 M^4} e^{-s(u)/M^2}f(u)+\Delta(c(u_0,s_0)), \label{Theta2}
\end{eqnarray}
where
\begin{eqnarray}
\delta(c(u,s_0))&=& e^{-s_0/M^2}\frac{f(u_0)}{{\cal C}_0}, \nn\\
\Delta(c(u,s_0))&=& e^{-s_0/M^2}\bigg[\frac{1}{2 u_0 M^2}\frac{f(u_0)} {{\cal C}_0} \nn\\
&&\left. -\frac{u_0^2}{2 {\cal C}_0} \frac{d}{du}\left( \frac{f(u)}{u{\cal C}} \right) \right|_{u = {u_0}}\bigg], \nn
\end{eqnarray}
${\mathcal C}_0 = m_b^2 + {u_0^2}m_\rho ^2 - {q^2}$ and $u_0$ is the solution of $c(u_0,s_0)=0$ with $0\leq u_0\leq 1$. Here we do not present the surface terms for the 3-particle DAs, whose contributions are quite small and can be safely neglected.

As a check of the above LCSRs, Eqs.(\ref{TFF_A1},\ref{TFF_A2},\ref{TFF_V}), it is found that by taking only the leading-twist terms in those LCSRs, we return to the LCSRs of Ref.\cite{LiZH09:Heavylight}.

It is noted that the LCSRs, Eqs.(\ref{TFF_A1},\ref{TFF_A2},\ref{TFF_V}), only contain the chiral-odd DAs' contributions. Then, we can use a more simpler way to get the LCSRs for those TFFs. That is, we can deal with the correlator, Eq.(\ref{correlator:1}), within the space-like region by directly treating their Dirac structures. The time-order production of this correlator, i.e. ${\rm T} \{\bar q_1(x)\gamma_\mu(1-\gamma_5) b(x), j_B^\dag (0)\} $, can be rewritten in a trace form as
\begin{displaymath}
{\rm Tr}\left\{[\underbrace{q_1(\bar u p)q_2(up)}_{\rm wavefunction}]\gamma_\mu(1-\gamma_5)(u\not\!p + \not\!q +m_b)(1+\gamma_5)\right\},
\end{displaymath}
in which the chiral-odd vacuum-to-vector state matrix element can be expanded as
\begin{eqnarray}
&&\langle \rho (p,\lambda )|{{\bar q}_{1a}}(x){q_{2b}}(0)|0\rangle_{x^2\to0}^{\rm chiral-odd}  = \frac{1}{4}\int_0^1 d u{e^{iup\cdot x}}\nn\\
&&~~~~\times \bigg\{ - if_\rho ^ \bot {\sigma _{\alpha \beta }}\bigg[{e^{*(\lambda )\alpha }}{p^\beta }\left[\phi _{2;\rho }^ \bot (u) + \frac{1}{{16}}m_\rho ^2{x^2}\phi _{4;\rho }^ \bot (u)\right]\nn\\
&&~~~~- m_\rho ^2{p^\alpha }{x^\beta }({e^{*(\lambda )}}\cdot x){I_L}(u) - \frac{1}{2}i m_\rho ^2{e^{*(\lambda )\alpha }}{x^\beta }{H_3}(u)\bigg]\nn\\
&&~~~~-\frac{i}{2}m_\rho^2f_\rho^\bot (e^{*(\lambda)}\cdot x) \psi_{3;\rho}^\|(u) \bigg\}_{ba}.  \nonumber
\end{eqnarray}
Using this trace form, we can get the same LCSRs for $B\to\rho$ TFFs.

\section{Numerical Analysis} \label{sec:III}

\subsection{Input parameters}

\begin{table}[htb]
\centering
\begin{tabular}{c c c c }
\hline
~~$m_b/{\rm GeV}$~~ & ~~$s_0/{\rm GeV}^2$~~ & ~~$M^2/{\rm GeV}^2$~~ & ~~$f_B/{\rm GeV}$~~ \\
\hline
$4.75$ & $[33.1, 36.9]$ & $[2.09, 2.57]$ & $0.179(5)$\\
\hline
$4.80$ & $[32.8, 35.9]$ & $[1.93, 2.36]$ & $0.160(5)$ \\
\hline
$4.85$ & $[32.5, 34.9]$ & $[1.81, 2.17]$ & $0.141(4)$ \\
\hline
\end{tabular}
\caption{A LCSR estimation on $f_B$ for $m_b=4.80\pm0.05$ GeV. The number in the parenthesis shows the uncertainty in the last digit.} \label{fB_Tab}
\end{table}

In doing the numerical calculation, we take $f_\rho^\bot=0.165(9)~{\rm GeV}$~\cite{Ball1998:Brho} and $m_b=4.80\pm0.05\;{\rm GeV}$ for the $b$-quark pole mass. The $\rho$-meson and $B$-meson masses are taken as ${m_\rho} = 0.775$ GeV and ${m_B} = 5.279$ GeV~\cite{pdg}. The value of $f_B$ can be consistently determined from a chiral LCSR, following the formulas in Ref.\cite{chiral}, we recalculate it and put the numerical results in Table \ref{fB_Tab}. Here, the Borel parameter $M^2$ and $s_0$ are determined by the requirements of the continuum state's contribution to be less than $30\%$ and the six condensates' contributions to be less than $10\%$ of the total LCSR.

\subsubsection{Models for the twist-2 LCDA $\phi_{2;\rho}^\bot$}

As shown by the LCSRs, Eqs.(\ref{TFF_A1},\ref{TFF_A2},\ref{TFF_V}), the $\rho$-meson LCDAs $\phi_{2;\rho}^\|$, $\phi_{3;\rho}^\bot$, $\psi_{3;\rho}^\bot$ and $\Phi_{3;\rho}^\|,\wt\Phi_{3;\rho }^\|$, which are at the $\delta^1$-order, provide zero contributions. The dominant contribution comes from the leading-twist LCDA $\phi_{2;\rho}^\bot$, while all the remaining twist-3 and twist-4 LCDAs contribute totally less than $10\%$ to the LCSRs. Thus, the uncertainties of the LCSR from the uncertainties of those high-twist LCDAs are highly suppressed. For clarity, we take those high-twist LCDAs directly as the ones suggested by Ref.\cite{Ball07:rhoWF}, which are put in Appendix \ref{sec:highertwist}.

The LCSRs, Eq.(\ref{TFF_A1},\ref{TFF_A2},\ref{TFF_V}), provide good platform for testing the properties of the leading-twist DA $\phi_{2;\rho}^\bot$. For the purpose, we adopt three models for $\phi_{2;\rho}^\bot$ to do the calculation, i.e. the Gegenbauer polynomial expansion with specific Gegenbauer moments, the one from the AdS/QCD theory, and the one from the Wu-Huang prescription, respectively.

Conventionally, one can expand the light meson's LCDA as a Gegenbauer expansion. As for $\phi_{2;\rho}^\bot$, it can be expanded as
\begin{eqnarray}
\phi_{2;\rho}^\bot(x,\mu_0)&=&6 x\bar x \left[1 + \sum\limits_{n=2,4,\ldots} a_{n}^\bot(\mu_0)  C^{3/2}_{n}(2x-1)\right],
\end{eqnarray}
where $\mu_0$ stands for some hadronic scale $\sim 1$ GeV. $a_{n}^\bot$ stands for $n_{\rm th}$-Gegenbauer moment and $C^{3/2}_{n}$ is the Gegenbauer polynomials. Such a Gegenbauer expansion is convergent and is dominated by its first several moments. Practically, one always takes the first term to do the analysis, i.e.
\begin{eqnarray}
\phi\rhob(x,\mu_0)=6x\bar x\bigg[1 + a_2^\bot(\mu_0) C_2^{3/2}(2x-1)\bigg]. \label{gegenbauer_1}
\end{eqnarray}
Several values for the second Gegenbauer moment $a_2^\bot(\mu_0=1\;{\rm GeV})$ from the QCD sum rules have been suggested in the literature, e.g. the one suggested by Chernyak and Zhitnitsky is $-0.167$~\cite{CZlike:Report} (we call it the CZ model) and the one suggested by Ball and Braun is $0.14\pm0.06$~\cite{Ball07:rhoWF} (we call it the BB model). It is noted that the CZ model prefers a single-peak behavior, while BB model tends to a double-humped behavior. When changing $a_2^\bot(\mu_0)$ from $0.08$ to $0.20$, the BB $\phi_{2;\rho}^\bot$ quickly varies from a flat single-peak behavior to a double-humped behavior. In the following, we shall take $a_2^\bot(\mu_0=1\;{\rm GeV})=0.14$ as the typical value for BB model. Furthermore, the Gegenbauer moments $a_n^\bot$ at any other renormalization scale can be obtained from QCD evolution, e.g. $a_n^\lambda(\mu)= a_n^\lambda(\mu_0){\cal L}^{\gamma_{n}^\lambda/\beta_0}$~\cite{Ball1998:Brho}, where $\beta_0=11-2n_f/3$, $\mathcal{L}=\alpha_s(\mu)/\alpha_s(\mu_0)$ and one-loop anomalous dimensions $\gamma^\bot_{\rho}=4C_F[\psi(n+2)+\gamma_E-1]$ with $\psi(n+1)=\sum_{k=1}^n 1/k -\gamma_E$. When the scale tends to infinity, we shall obtain $a_n^{\lambda}(\infty)\to0$, which corresponds to the asymptotic DA suggested by Ref.\cite{TFFas}.

On the other hand, the $\rho$-meson DA $\phi_{2;\rho}^\bot$ can be derived from its LCWF, since it can be related with the LCWF via the relation
\begin{eqnarray}
\phi_{2;\rho}^\bot(x,\mu_0) = \frac{ 2\sqrt{3}}{ \wt{f}_\rho^\bot}\int_{|{\bf k}_\bot|^2 \leq\mu^2_0}\frac{d{\bf k}_\bot}{16\pi^3}\psi_{2;\rho}^\bot(x,{\bf k}_\bot), \label{DA_WF}
\end{eqnarray}
where $\wt{f}_\rho^\bot = f_\rho^{\bot}/{\cal C}_\rho^\bot $ is the improved vector decay constant with ${\cal C}_\rho^\bot=\sqrt{3}$.

One way of constructing the LCWF has been suggested under the AdS/QCD theory~\cite{Brodsky:2007hb,Brodsky:2008pg}. That is, Ref.\cite{AdS:2} suggests
\begin{eqnarray}
\psi_{2;\rho}^\bot(x,\zeta)= \mathcal{N}_{\bot} \frac{\kappa} {\sqrt{\pi}}\sqrt{x\bar{x}} \exp \left(-\frac{\kappa^2 \zeta^2}{2}\right) \exp\left(-\frac{m_f^2}{2\kappa^2 x\bar{x}} \right),\nn\\
\end{eqnarray}
which leads to
\begin{eqnarray}
\phi_{2;\rho}^{\bot}(x,\mu_0) &=&\frac{ 3 m_f}{\pi f_\rho^\bot} \int d\zeta \mu_0 J_1(\mu_0 \zeta) \frac{\psi_{2;\rho}^\bot(x,\zeta)}{x\bar x}. \label{phi_4}
\end{eqnarray}
We call it the AdS/QCD model. The parameter $m_f=0.14{\rm GeV}$~\cite{AdS:1,AdS:2,Soyez:2007kg, Forshaw:2006np, Forshaw:2004vv} and ${\cal N}_\bot= 2.031$, which is fixed by the normalization condition, $\int_0^1 \phi_{2;\rho}^{\bot({\rm AdS})}(x,\mu_0) = 1$. $\kappa^2 = m^2_\rho/2$ and $\zeta= r \sqrt{x\bar{x}}$ with $r$ being the transverse distance between the quark and antiquark at the equal light-front time and $\sqrt{x\bar x}$ being the variable that maps onto the fifth dimension of the AdS space \cite{Brodsky:2012je,Brodsky:2013je,deTeramond:2012cs}.

Another way of constructing the light-meson WF has been suggested by Wu and Huang~\cite{XGWu:2010} (we call it the WH model). Following its idea, the radial part $\psi _{2;\rho}^{R}$ of $\phi_{2;\rho}^\bot$ can be constructed from the BHL-prescription~\cite{BHL} and its spin-space part $\chi_\rho^{h_1 h_2}(x,\bf{k}_\bot)$ can be derived from the Wigner-Melosh rotation~\cite{WF94,WFspin}, that is
\begin{eqnarray}
\psi_{2;\rho}^\bot(x,{{\bf{k}}_ \bot }) = \sum\limits_{{h _1}{h _2}} {{\chi_\rho^{{h _1}{h_2}}}} (x,{{\bf{k}}_ \bot }) \psi _{2;\rho}^{R}(x,{{\bf{k}}_ \bot }),
\end{eqnarray}
where $\psi _{2;\rho}^{R} \propto [1 + {B_{2;\rho}^\bot }C_2^{3/2}(\xi )] \exp \left[-b_{2;\rho}^{\bot 2}\frac{\bfk^2 + m_q^2}{x\bar x}\right]$ and the spin-space wavefunction ${{\chi_\rho^{{h _1}{h_2}}}} (x,{{\bf{k}}_ \bot })$ can be found in Ref.\cite{BHL}. Then, we get
\begin{widetext}
\begin{eqnarray}
&&\phi _{2;\rho }^\bot (x,\mu_0) = \frac{{A_{2;\rho}^\bot \sqrt {3x\bar x} {m_q}}}{{8{\pi ^{3/2}}\wt f_\rho ^\bot b_{2;\rho}^\bot }}[1 + {B_{2;\rho}^\bot }C_2^{3/2}(\xi )] \times \left[ {{\rm{Erf}}\left( {b_{2;\rho}^\bot \sqrt {\frac{{{\mu^2_0} + m_q^2}}{{x\bar x}}} } \right) - {\rm{Erf}}\left( {b_{2;\rho}^\bot \sqrt {\frac{{m_q^2}}{{x\bar x}}} } \right)} \right], \label{DA:WH}
\end{eqnarray}
\end{widetext}
where the error function $\textrm{Erf}(x) =\frac{2}{\sqrt{\pi}}\int^x_0 e^{-t^2} dt$ and the constitute quark mass $m_q\simeq300$ MeV. In addition to the normalization condition, we adopt the average value of the squared transverse momentum $\langle{\bf k}_\bot^2\rangle_{2;\rho}$ as another constraint, which is defined as
\begin{displaymath}
\langle{\bf k}_\bot^2\rangle_{2;\rho}=\frac{\int dx d^2{\bf k}_\bot |{\bf k}_\bot|^2 |\psi_{2;\rho}^\bot(x,{\bf k}_\bot)|^2}{\int dx d^2 {\bf k}_\bot|\psi_{2;\rho}^\bot(x,{\bf k}_\bot)|^2}.
\end{displaymath}
We take the value of $\langle{\bf k}_\bot^2\rangle^{1/2}_{2;\rho}$ to be $0.37$ GeV, which is consistent with the choice of Refs.\cite{XHGuo1991,XGWu:2010} for the light-mesons. The Gegenbauer moments can be derived from the equation
\begin{equation}
a_n^\bot(\mu_0)=\frac{\int_0^1 dx~ \phi_{2;\rho}^\bot(x,\mu_0) C_n^{3/2}(2x-1)}{\int_0^1 dx~ 6x \bar x[C_n^{3/2}(2x-1)]^2}.
\end{equation}
Using this equation, we can obtain a relation between $B\rhob$ and $a_2^\bot$.

\begin{table}[htb]
\centering
\begin{tabular}{c c c c c}
\hline
~$B_{2;\rho}^\bot$~    & ~$A_{2;\rho}^\bot({\rm GeV^{-1}})$~ & ~$b_{2;\rho}^\bot({\rm GeV}^{-1})$~ &  ~$a_2^{\bot}(1{\rm GeV})$~ &  ~$a_2^{\bot}(2.2{\rm GeV})$~ \\ \hline
  $-0.2$	&	 28.56 	&	0.643 	&	 $-0.180$    &	$-0.152$\\
  $-0.1$	&	 27.50 	&	0.628 	&	 $-0.080$    &	$-0.067$\\
  $	0.0$	&	 25.88 	&	0.604 	&	 $+0.026$	 &  $-0.022$\\
  $+0.1$	&	 23.82 	&	0.572 	&	 $+0.140$	 &  $+0.118$\\
  $+0.2$	&	 21.61 	&	0.537 	&	 $+0.258$	 &  $+0.217$\\
\hline
\end{tabular}
\caption{The $\rho$-meson leading-twist LCDA parameters $A\rhob$ and $b\rhob$ for some typical choices of $B_{2;\rho}^\bot$. The resultant second Gegenbauer moment $a_2^{\bot}$ at $\mu_0=1.0{\rm GeV}$ and $2.2{\rm GeV}$ are also presented. }   \label{DA_parameter_1}
\end{table}

\begin{figure}[htb]
\includegraphics[width=0.45\textwidth]{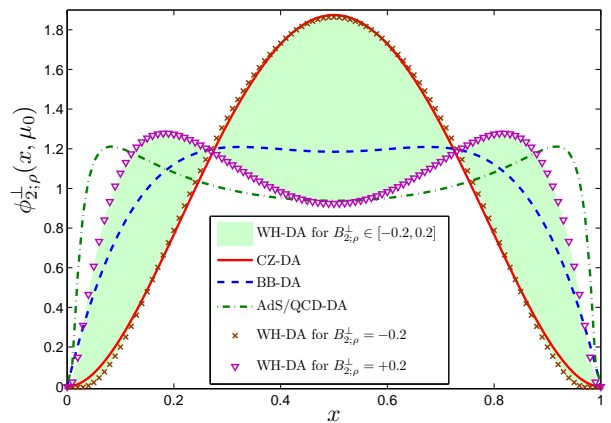}
\caption{A comparison of $\phi_{2;\rho}^\bot(x,\mu_0=1{\rm GeV})$ under various models. As for WH-DA model, which are shown by shaded band, we adopt $B_{2;\rho}^{\bot}\in[-0.2,0.2]$. }\label{DA:2bot}
\end{figure}

We put the $\rho$-meson leading-twist DA parameters in Table \ref{DA_parameter_1}, where $B_{2;\rho}^{\bot}\in[-0.2,+0.2]$. The resultant second Gegenbauer moment $a_2^{\bot}$ at $\mu_0=1.0{\rm GeV}$ and $2.2{\rm GeV}$ are also presented. We observe that $B_{2;\rho}^{\bot}\sim a_2^{\bot}$, which indicates that $B_{2;\rho}^\bot$ dominantly determines the longitudinal distribution. We present a comparison of $\phi_{2;\rho}^\bot(x,\mu_0)$ under various models in Fig.(\ref{DA:2bot}). When $B_{2;\rho}^\bot$ changes from $-0.20$ to $+0.20$, the $\rho$-meson DA varies from the single-peak behavior to the double-humped behavior. More specifically, we find that $B_{2;\rho}^\bot=-0.20$ corresponds to CZ-DA, $B_{2;\rho}^\bot=0.10$ corresponds to BB-DA and $B_{2;\rho}^\bot=0.32$ corresponds to AdS/QCD-DA. Thus the WH-DA provides a convenient form to mimic the behavior of various DA models suggested in the literature. If we have precise measurements for certain processes involving $\rho$-meson, then by comparing the theoretical estimations, we can fix the value of $B_{2;\rho}^\bot$ and then get the $\rho$-meson DA's behavior. As will be shown later that the WH-DA with $B_{2;\rho}^{\bot}>0.10$ leads to a small $A_1(q^2)$ in comparison to the lattice QCD estimation, so we shall take $B_{2;\rho}^{\bot}\in[-0.20, +0.10]$ to do our following discussions.

\subsubsection{LCSR parameters for $B\to\rho$ TFFs}

\begin{table}[htb]
\centering
\begin{tabular}{ l c c c c}
\hline
& &WH-DA& & \\
&$B\rhob=0.0$&$B\rhob=0.1$&$B\rhob=-0.2$& AdS/QCD-DA \\\cline{2-4}
&$ / $& ~BB-like DA~ & ~CZ-like DA~ & \\
\hline
$s_0^{A_1}$ &35.5(5)&33.4(6)&38.0(5)&31.0(5)\\
$M^2_{A_1}$ &7.4(5) &7.2(5) &7.2(5) &7.1(5)\\
\hline
$s_0^{A_2}$ &35.9(5)&32.9(5)&38.4(5)&31.0(5)\\
$M^2_{A_2}$ &7.0(5) &7.1(5) &8.1(5) &6.8(5)\\
\hline
$s_0^{V}$ &36.5(5)&34.5(5)&39.0(5)&30.8(5)\\
$M^2_V$   &10.5(5)&10.0(5)&9.9(5) &7.9(5)\\
\hline
\end{tabular}
\caption{The determined continuum threshold $s_0$ and the Borel parameter $M^2$ under the WH-model for $B\rhob=-0.2,0.0,+0.10$, respectively. As a comparison, the results for the AdS/QCD-DA are also presented. The central values are for $m_b = 4.80$ GeV and $f_B = 0.160$ GeV.}  \label{s0M2}
\end{table}

We adopt the following conditions to set the values for the LCSR parameters, such as the Borel window $M^2$ and the continuum threshold $s_0$, of the $B\to\rho$ TFFs. Firstly, we require the continuum contribution to be less than $30\%$ of the total LCSR, i.e.
\begin{eqnarray}
\frac{\int_{s_0}^\infty ds \rho^{\rm tot} (s) \exp \big[- s/M^2\big]}{\int_{m_b}^\infty ds \rho^{\rm tot}(s) \exp \big[-s/M^2\big]}< 30\%,
\end{eqnarray}
which can be further translated as
\begin{eqnarray}
\frac{\int_0^{u_0}{du} \rho^{\rm tot}(u)\exp\big[-s(u)/M^2\big]}{ \int_0^1 du \rho^{\rm tot}(u)\exp\big[-s(u)/M^2\big] } < 30\%.
\end{eqnarray}
The spectral density $\rho^{\rm tot}$ for specific $B\to\rho$ TFFs can be read from Eqs.(\ref{TFF_A1},\ref{TFF_A2},\ref{TFF_V}). Secondly, we require all the higher-twist DAs' contributions to be less than $15\%$ of the total LCSR. The derivative of Eqs.(\ref{TFF_A1},\ref{TFF_A2},\ref{TFF_V}) with respect to $-1/M^2$ gives the LCSR for the $B$-meson mass $m_B$. Then, as a third constraint, we require the estimated $B$-meson mass to be fulfilled with high accuracy $\sim0.1\%$ in comparing with the experiment one, e.g. ${|m_B^{\rm SR}- m_B^{\rm exp}|}/{m_B^{\rm exp}}<0.1\%$. As for the continuum threshold $s_0$, it is usually set to be close to the squared mass of the $B$-meson's first exciting state. We set its value to be within a broader range, $s_0\in[30,39]{\rm GeV}^2$. We present the values of $s_0$ and $M^2$ for the $B\to\rho$ TFFs in Table \ref{s0M2}, where as a comparison, the results for the AdS/QCD-DA are also presented.

\subsection{The $B\to\rho$ TFFs and the CKM matrix element $\Vub$}

\begin{table*}[tb]
\begin{tabular}{c  c c c  | c c c | c c c  }
\hline
    &\multicolumn{3}{c|}{$B\rhob=-0.2$}&\multicolumn{3}{c|} {$B\rhob=0.0$}&\multicolumn{3}{c}{$B\rhob=0.1$}\\ \cline{2-10}
&~~~~$A_1(0)$~~~~&~~~~$A_2(0)$~~~~&~~~~$V(0)$~~~~&~~~~$A_1(0)$~~~~&~~~~$A_2(0)$~~~~&~~~~$V(0)$~~~~&~~~~$A_1(0)$~~~~&~~~~$A_2(0)$~~~~&~~~~$V(0)$~~~~ \\
\hline
$	\phi_{2;\rho}^\bot	$&$	0.214	$&$	0.267 	$&$	0.301	$&$	0.244	$&$	 0.305 	$&$	0.340	$&$	 0.245	 $&$	0.306 	$&$	0.343	$	\\
$	\psi_{3;\rho}^\|	$&$	0.007	$&$	-0.036 	$&$	      /	$&$	0.006	$&$	 -0.040 	$&$	      /	$&$	 0.006	$&$	-0.040 	$&$	      /	$	\\
$	\phi_{4;\rho}^\bot 	$&$	-0.022	$&$	-0.035 	$&$	-0.031	$&$	-0.022	$&$	 -0.039 	$&$	-0.031	$&$	 -0.023	$&$	-0.040 	$&$	-0.031	$	\\
$	I_L             	$&$	-0.003	$&$	-0.022 	$&$	      /	$&$	-0.003	$&$	 -0.022 	$&$	      /	$&$	 -0.003	$&$	-0.022 	$&$	      /	$	\\
$	H_3             	$&$	0.009	$&$	0.005 	$&$	      /	$&$	0.006	$&$	 0.004 	$&$	      /	$&$	 0.005	 $&$	0.003 	$&$	      /	$	\\
$	\Phi_{3;\rho}^\bot	$&$	      /	$&$	0.0004 	$&$	      /	$&$	      /	$&$	 0.0004	$&$	      /	$&$	      /	$&$	0.0003	$&$	      /	$	\\
$	\rm{Total}        	$&$	0.204	$&$	0.179 	$&$	0.270	$&$	0.232	$&$	 0.208 	$&$	0.309	$&$	 0.231	 $&$	0.207 	$&$	0.311	$	\\ \hline
\end{tabular}
\caption{The $B\to\rho$ TFFs at the large recoil region, $q^2=0$, in which the twist-2, the non-zero twist-3 and twist-4 DAs' contributions are presented separately. The WH-DA has been adopted in the calculation, in which $B\rhob=0.0$, $0.1$ and $-0.2$, respectively. The scale $\mu$ is set as $2.2{\rm GeV}$. } \label{Endingpoint_1}
\end{table*}

We present the $B\to\rho$ TFFs at the large recoil region, $q^2=0$, in Table \ref{Endingpoint_1}, in which the contributions from the DAs with various twist structures are presented. Numerically, we have found that the twist-3 DA $\phi_{3;\rho}^\|$ and twist-4 DA $\psi_{4;\rho}^{\bot}$ contain large twist-2 component $\phi_{2;\rho}^\bot$, then $\phi_{3;\rho}^\|$ and $\psi_{4;\rho}^{\bot}$ shall have large contributions to the LCSR, which does not satisfy the twist-power counting. However, Table \ref{Endingpoint_1} shows that $I_L$ and $H_3$ follow the $\delta$-power counting. Thus, it is better to use the $\delta$-power counting other than the usual twist-powering counting to deal with their contributions. This observation has already been found in Ref.\cite{Ball04:Brho}. The 3-particle high-twist DAs' contributions such as those of $\Phi_{3;\rho}^\bot$ are only about $0.1\%$ of the total LCSR. For example, the non-zero contribution from the 3-particle DA $\Phi_{3;\rho}^\bot$ is only about $-0.2\%$ to $A_2(0)$. Then, those 3-particle DAs' contributions can be safely neglected.

\begin{table}[tb]
\centering
\begin{tabular}{ c  c  c  c }
\hline
&~~$A_1(0)$~~&~~$A_2(0)$~~&~~$V(0)$  \\  \hline
WH($B\rhob=-0.2$) & $0.204^{+0.011}_{-0.011}$ & $0.179^{+0.012}_{-0.014}$ & $0.270^{+0.012}_{-0.012}$ \\
WH($B\rhob=~~0.0$) & $0.232^{+0.010}_{-0.010}$ & $0.208^{+0.012}_{-0.013}$ & $0.309^{+0.014}_{-0.014}$ \\
WH($B\rhob=+0.1$) & $0.231^{+0.010}_{-0.010}$ & $0.207^{+0.011}_{-0.012}$ & $0.311^{+0.014}_{-0.012}$ \\
BB-DA & $0.234^{+0.010}_{-0.010}$ & $0.214^{+0.011}_{-0.012}$ & $0.315^{+0.012}_{-0.014}$ \\
CZ-DA & $0.218^{+0.011}_{-0.011}$ & $0.200^{+0.012}_{-0.013}$ & $0.289^{+0.011}_{-0.013}$ \\
AdS/QCD-DA & $0.277^{+0.011}_{-0.014}$ & $0.257^{+0.014}_{-0.015}$ & $0.347^{+0.013}_{-0.015}$ \\
\hline
\end{tabular}
\caption{The $B\to\rho$ TFFs at the large recoil region under the WH-DA, BB-DA, CZ-DA and AdS/QCD-DA, where the errors are squared average of all the mentioned sources. The scale $\mu$ is set as $2.2{\rm GeV}$. } \label{Endingpoint_2}
\end{table}

The $\rho$-meson leading-twist DA $\phi_{2;\rho}^\bot$ provides the dominant contribution to $B\to\rho$ TFFs. To show how $\phi_{2;\rho}^\bot$ affects the TFFs, we present the $B\to\rho$ TFFs at the large recoil region under the WH-DA, BB-DA, CZ-DA and AdS/QCD-DA in Table \ref{Endingpoint_2}, where the errors are squared average of the uncertainties from the above mentioned input parameters. In doing the calculation, the scale $\mu$ is set as the typical energy of the process, i.e. $\mu\simeq\sqrt{m_B^2-m_b^2}\sim 2.2{\rm GeV}$. As for WH-DA, we calculate the cases with $B\rhob=-0.2$, $0.0$ and $0.1$, respectively. The results for $B\rhob=-0.2$ are close to that of CZ-DA and the results for $B\rhob=0.1$ are close to that of BB-DA, which are due to their close DA behaviors as shown by Fig.(\ref{DA:2bot}). Table \ref{Endingpoint_2} shows that all the TFFs increases with the increment of $B\rhob$, i.e. a larger TFFs can be achieved for a larger $B\rhob$, or equivalently, a larger second Gegenbauer moment $a^{\perp}_2$. As a further cross check of the present LCSR, if taking the same second Gegenbauer moment $a^{\perp}_2$ as that of Ref.\cite{Ball1998:Brho,Ball04:Brho}, we find that our present TFFs agree with those derived under the usual choice of correlator~\cite{Ball1998:Brho,Ball04:Brho}. Moreover, our present results for the TFFs at $q^2=0$ agree with the pQCD prediction~\cite{LuCD:2002}: $A_1(0)=0.25\pm0.02$, $A_2(0)=0.21\pm0.015$ and $V(0)=0.318\pm0.032$.

The LCSRs for $B\to\rho$ TFFs are valid when $\rho$-meson's energy ($E_\rho$) is large enough, i.e. $E_\rho\gg\Lambda_{\rm QCD}$, which implies a restriction for not too large $q^2$. Because $q^2=m_B^2-2m_B E_\rho$, we adopt $0\leq q^2\leq q^2_{\rm LCSR;MAX} \sim 14{\rm GeV^2}$ to evaluate the sum rules. As a comparison, it is noted that the maximum physical allowable value for $q^2$ is $(m_B-m_\rho)^2 \sim 20{\rm GeV^2}$. On the other hand, because of the restriction to $\rho$ energies which should be smaller than the inverse lattice spacing, the lattice QCD calculation becomes more difficult in large recoil region. At present, the lattice QCD results of $B\to\rho$ TFFs are available only for soft region, i.e., $q^2 > 12 {\rm GeV}^{2}$~\cite{Lattice96:1,Lattice96:2,Lattice04}. Thus, to compare the LCSR estimations with the lattice ones, certain extrapolation has to be done.

\begin{table}[htb]
\centering
\begin{tabular}{c c c c c }
\hline
 & ~~$F_i$~~ & ~~$a_i$~~ & ~~$b_i$~~ & ~~$\Delta$~~ \\
 \hline
              & $A_1$ & 1.351 & 0.682 & 0.2 \\ \cline{2-5}
$B\rhob=-0.2$ & $A_2$ & 2.159 & 1.430 & 0.4   \\ \cline{2-5}
              & $V$ & 2.041 & 1.228 & 0.8 \\
\hline
             & $A_1$ & 0.744 & 0.077 & 0.0 \\ \cline{2-5}
$B\rhob=0.0$ & $A_2$ & 1.566 & 0.734 & 0.2   \\ \cline{2-5}
             & $V$ & 1.580 & 0.619 & 0.1 \\
\hline
             & $A_1$ & 0.536 & 0.003 & 0.2 \\ \cline{2-5}
$B\rhob=0.1$ & $A_2$ & 1.405 & 0.654 & 0.3   \\ \cline{2-5}
             & $V$ & 1.403 & 0.439 & 0.2 \\
\hline
\end{tabular}
\caption{The fitted parameters $a_i$ and $b_i$ defined in Eq.(\ref{fit}) for the $B\to\rho$ TFFs with all the LCSR parameters set to be their central values. $\Delta$ is a measure of the quality of the fit defined in Eq.(\ref{delta}).} \label{analytic}
\end{table}

\begin{figure}[htb]
\includegraphics[width=0.45\textwidth]{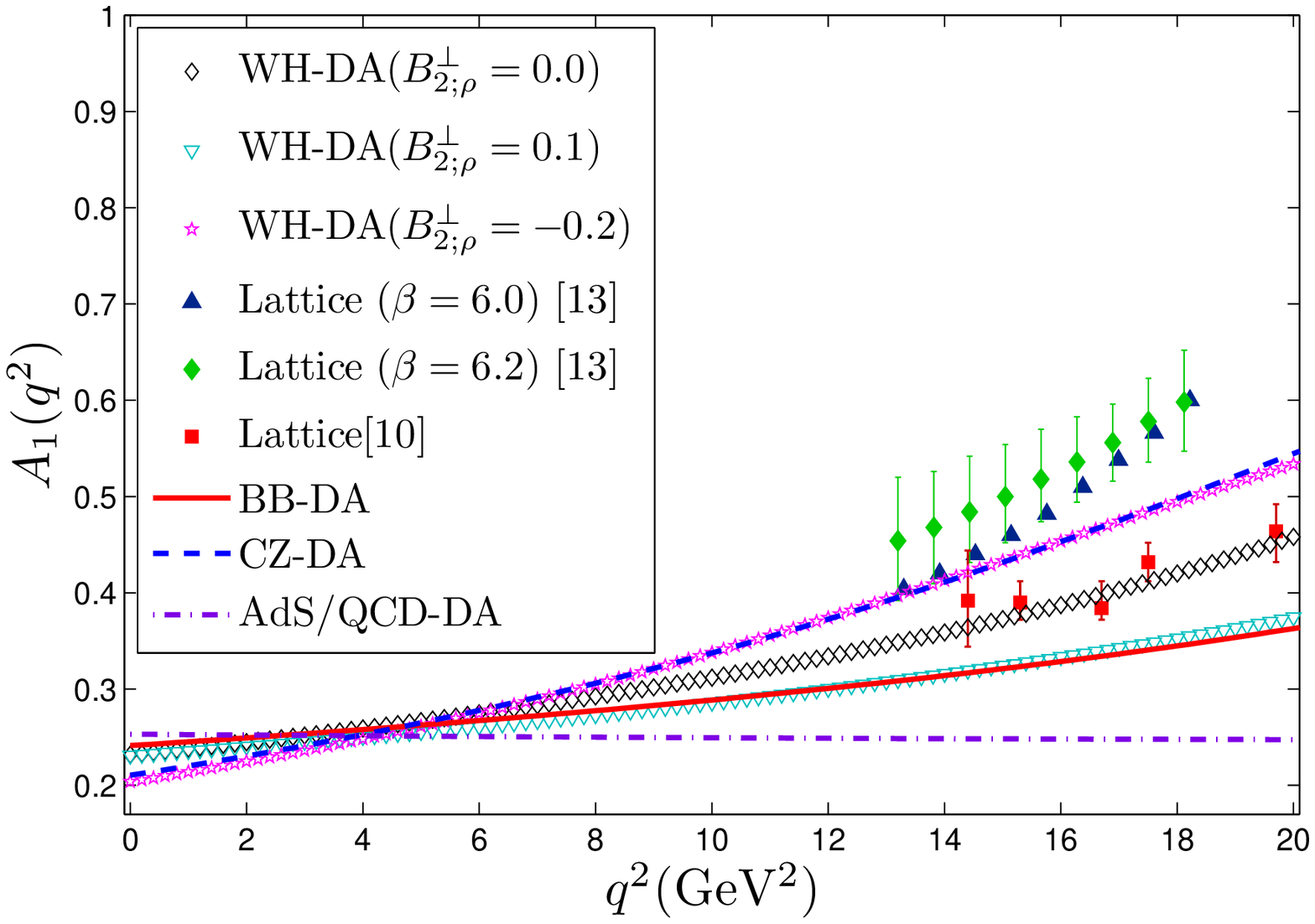}
\includegraphics[width=0.45\textwidth]{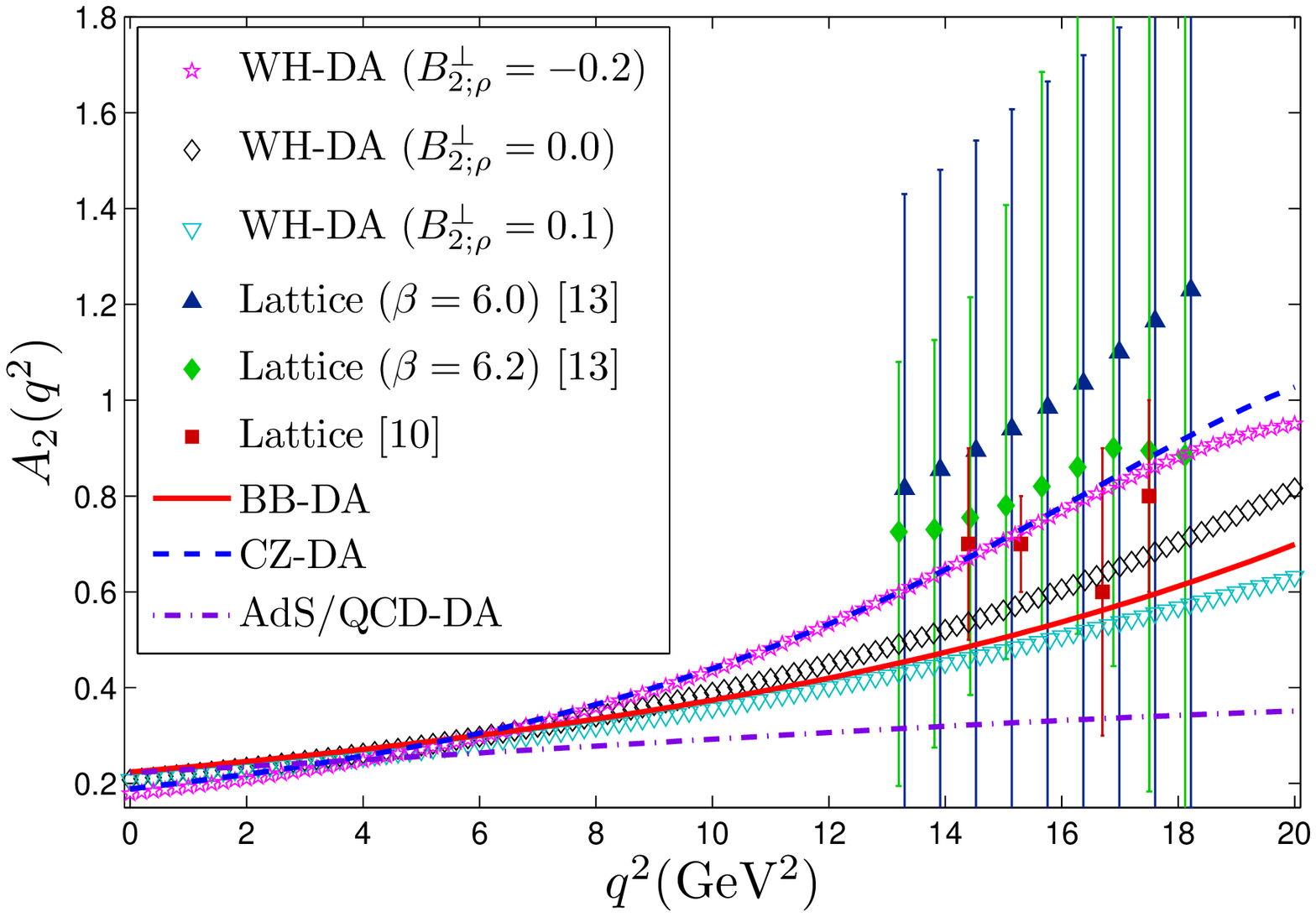}
\includegraphics[width=0.45\textwidth]{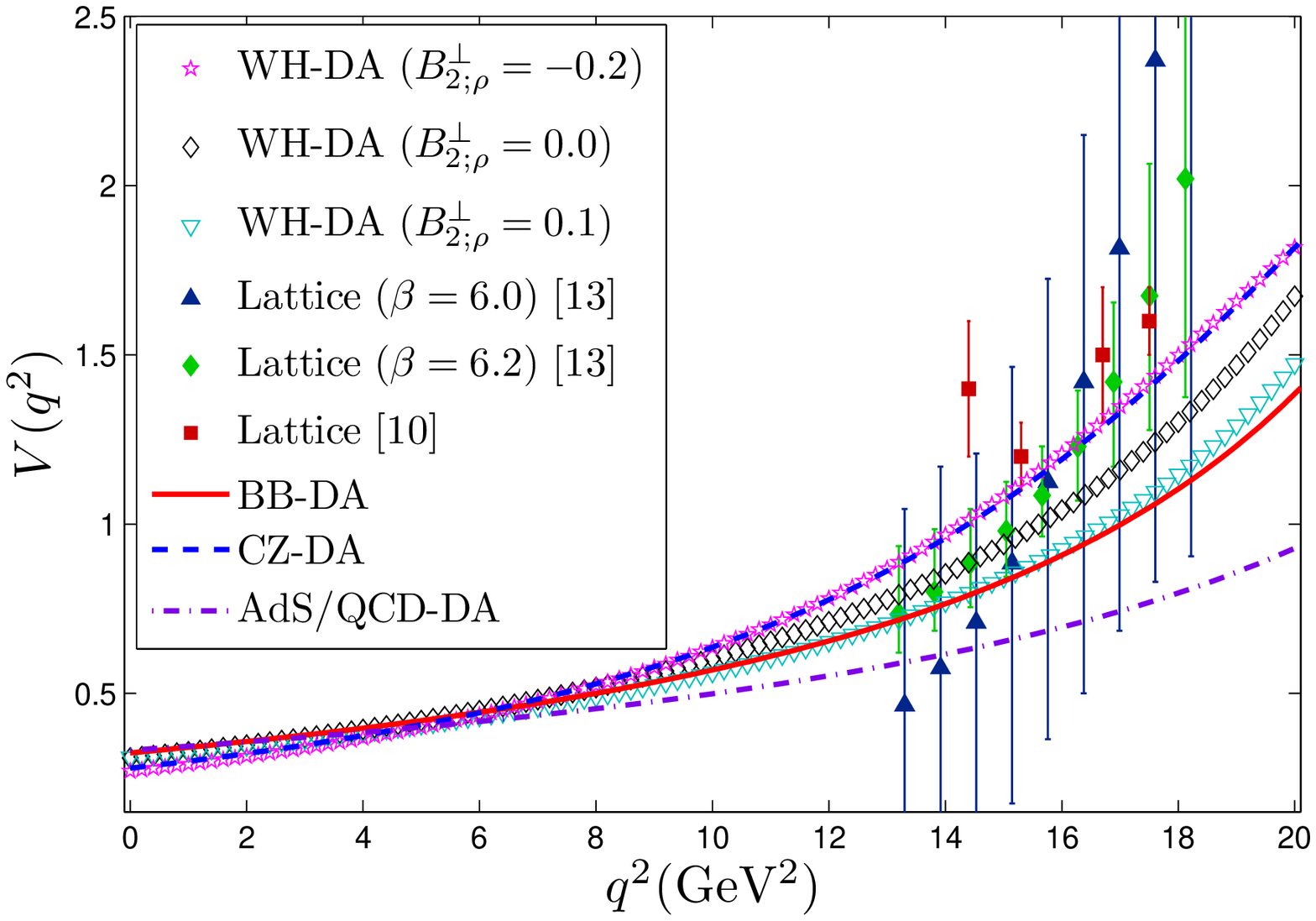}
\caption{The extrapolated $A_1(q^2)$, $A_2(q^2)$ and $V(q^2)$ under the WH-DA, BB-DA, CZ-DA and AdS/QCD-DA, respectively. The lattice QCD estimations~\cite{Lattice96:1,Lattice04} are included for a comparison. } \label{TFF:A1A2V}
\end{figure}

We take our present LCSR estimations within the region of $q^2\in[0,14]{\rm GeV}^2$ to do the extrapolation. We adopt the following formulae for the extrapolation,
\begin{equation}
F_i(q^2)=\frac{F_i(0)}{1-a_i q^2/m_B^2 + b_i(q^2/m_B^2)^2}, \label{fit}
\end{equation}
where $F_i$ stands for the mentioned $B\to\rho$ TFFs, i.e. $A_1$, $A_2$ and $V_1$, accordingly. When $b_i=0$, we return to the usual vector meson dominance extrapolation. The parameters $a_i$ and $b_i$ are fitted by requiring the ``quality'' of the fit to be within $1\%$, i.e. $\Delta<1$. Here, the ``quality" of the fit is expressed by a parameter $\Delta$, which is defined as~\cite{Ball05}
\begin{equation}
\Delta=100\frac{\sum_t\left|F_i(t)-F_i^{\rm fit}(t)\right|} {\sum_t\left|F_i(t)\right|}, \label{delta}
\end{equation}
where $t\in[0,\frac{1}{2},\cdots,\frac{27}{2},14]{\rm GeV}^2$. The fitted parameters are put in Table \ref{analytic}.

We put the extrapolated $B\to\rho$ TFFs $A_1(q^2)$, $A_2(q^2)$ and $V(q^2)$ in Fig.(\ref{TFF:A1A2V}), in which the lattice QCD estimations~\cite{Lattice96:1,Lattice04} are included as a comparison. The TFFs become smaller with the increment of $a^\perp_2$. This indicates that a larger second Gegenbauer moment $a^\perp_2$ is not allowed by the lattice QCD estimations. If we have a more precise lattice QCD estimation, we can get a more strong constraint on $\rho$-meson LCDA behavior.

As mentioned above, the WH-DA provides a convenient $\rho$-meson DA model for mimicking the behaviors of the LCDA models suggested in the literature, which are shown by Fig.(\ref{TFF:A1A2V}). For examples, the TFFs for WH-DA with $B^{\perp}_{2;\rho}\sim -0.2$ agree with the estimations of CZ-DA, and the TFFs for WH-DA with $B^{\perp}_{2;\rho}\sim0.1$ agree with the estimations of the BB-DA. If taking $B^{\perp}_{2;\rho}\sim0.3$ for WH-DA, we shall get the same prediction of AdS/QCD-model. Among all $\rho$-meson DA models, the AdS/QCD-model provide the smallest TFFs. It is noted that if taking a larger $m_f$ value as $0.35{\rm GeV}$~\cite{AdS2013:Brho}, corresponding to $a^\perp_2 \simeq 0.0$, we can obtain satisfactory results consistent with the lattice QCD estimations. In the following, we shall adopt WH-DA model for detailed discussions on $B\to\rho$ semi-leptonic decays.

\begin{figure}[bt]
\includegraphics[width=0.45\textwidth]{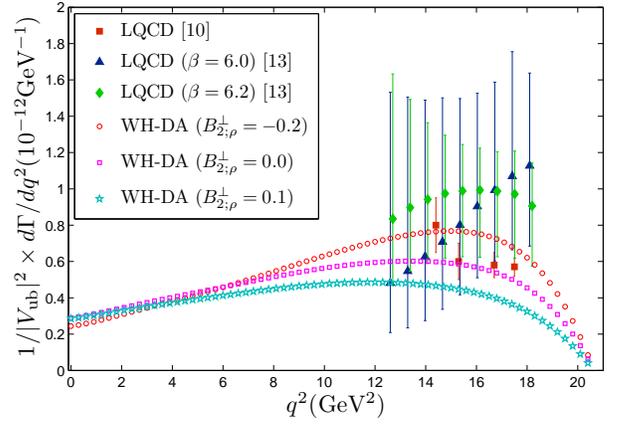}
\caption{The differential decay width $1/|V_{\rm ub}|^2 \times d\Gamma/dq^2$ for the WH-DA. The lattice QCD estimations~\cite{Lattice96:1,Lattice04} are included for a comparison. } \label{dGamma}
\end{figure}

We present the differential decay width $1/|V_{\rm ub}|^2 \times d\Gamma/dq^2$ for the WH-DA in Fig.(\ref{dGamma}), where $B\rhob$ is taken as $-0.2$, $0.0$ and $0.1$, respectively. The lattice QCD estimations~\cite{Lattice96:1,Lattice04} are included for a comparison.

\begin{table}[htb]
\begin{tabular}{cccc}
\hline
& $B\rhob=-0.2$ & $B\rhob=0.0$ & $B\rhob=0.1$   \\
\hline
${\Gamma}/{\Vub^2} ({\rm ps}^{-1})$ & $10.95^{+1.57}_{-1.39}$ &$9.57^{+1.34}_{-1.11}$ &$7.97^{+1.13}_{-0.97}$  \\
$\Gamma^{\|}/\Gamma^{\bot}$ &$0.79^{+0.15}_{-0.15}$ & $0.89^{+0.16}_{-0.15}$ & $0.93^{+0.17}_{-0.16}$  \\
\hline
\end{tabular}
\caption{Total decay width ${\Gamma}/{\Vub^2}$ and the ratio $\Gamma^\|/\Gamma^\bot$ under the WH-DA with $B\rhob=-0.2$, $0.0$ and $0.1$, respectively. The errors are squared average of the mentioned error sources. } \label{Gammatotal}
\end{table}

The total $B\to\rho$ semi-leptonic decay width can be separated as
\begin{equation}
\Gamma=\Gamma^\|+\Gamma^\bot, \label{Gamma}
\end{equation}
where $\Gamma^\|$ stands for the decay width of $\rho$-meson longitudinal components,
\begin{displaymath}
\Gamma^\|= {\cal G} |V_{\rm ub}|^2 \int_0^{q^2_{\rm max}} dq^2 \sqrt{\lambda(q^2)} q^2 H_0^2(q^2)
\end{displaymath}
and $\Gamma^\bot$ stands for the decay width of $\rho$-meson transverse components,
\begin{displaymath}
\Gamma^\bot={\cal G} |V_{\rm ub}|^2 \int_0^{q^2_{\rm max}} dq^2 \sqrt{\lambda(q^2)} q^2 [H_+^2(q^2)+H_-^2(q^2)].
\end{displaymath}
The total decay width ${\Gamma}$ and the ratio $\Gamma^\|/\Gamma^\bot$ computed for the WH-DA are presented in Table \ref{Gammatotal}, where the errors in Table \ref{Gammatotal} are squared average of the mentioned error sources. It is noted that the total decay width ${\Gamma}$ decreases and the ratio $\Gamma^\|/\Gamma^\bot$ increases with the increment of $B^{\perp}_{2;\rho}$ ($a^{\perp}_{2;\rho}$).

In the literature, the $B\to \rho$ semi-leptonic decays has also been adopted for determining the CKM matrix element $|V_{\rm ub}|$. Two types of semi-leptonic decays have been adopted for such purpose. The first type, the so-called ``$B^0$-type'', is via the process $B^0\to\rho^-\ell^+\nu_\ell$, whose branching ratio and lifetime are~\cite{pdg}
\begin{eqnarray}
{\cal B}(B^0\to\rho^-\ell^+\nu_\ell)&=&(2.34\pm 0.28)\times 10^{-4},\nn\\
\tau(B^0)&=& 1.519\pm 0.007 {\rm ps}.
\end{eqnarray}
The second type, the so-called ``$B^+$-type'', is via the process $B^+\to\rho^0\ell^+\nu_\ell$, whose branching ratio and lifetime are~\cite{pdg}
\begin{eqnarray}
{\cal B}(B^+\to\rho^0\ell^+\nu_\ell)&=& (1.07\pm 0.13)\times 10^{-4},\nn\\
\tau(B^+)&=& 1.641\pm 0.008 {\rm ps}.
\end{eqnarray}
The experimental measurements and the theoretical estimations can be related via the relation
\begin{eqnarray}
&&{\cal B}(B^0/B^+\to \rho^-/\rho^0)\tau(B^0/B^+)\nonumber\\
&=&\frac{{\cal G}\Vub^2} {c_\rho^2} \int_0^{q_{\rm max}^2} dq^2 \sqrt{\lambda(q^2)} q^2 \sum_{j=0,\pm} H_j^2(q^2),
\end{eqnarray}
where the factor $c_\rho$ accounts for $\rho$-meson flavor content, and we have
\[{c_\rho } = \left\{ {\begin{array}{*{20}{l}}
{\sqrt 2 \quad\quad {\rho ^0}(b \to u)}\\
{ - \sqrt 2 \quad {\rho ^0}(b \to d)}\;\;, \\
{ 1 \quad \quad {\rho ^ - }(b \to u,d)}
\end{array}} \right.\]
since $\rho^0 = (\bar uu-\bar dd)/\sqrt{2}$ and $\rho^{-}=\bar{u}d$.

\begin{table}[tb]
\centering
\begin{tabular}{ c c c }
\hline
   & ~~~$B^+$-type~~~ & ~~~$B^0$-type~~~ \\
\hline
~~$B\rhob=-0.2$~~ & ~~$2.80^{+0.19+0.17}_{-0.19-0.18}$~~ & ~~$3.04^{+0.20+0.18}_{-0.21-0.19}$~~
\\
$B\rhob=0.0$  & $2.99^{+0.18+0.18}_{-0.20-0.19}$ & $3.25^{+0.20+0.19}_{-0.22-0.20}$
\\
$B\rhob=+0.1$  & $3.28^{+0.21+0.19}_{-0.22-0.21} $ & $3.57^{+0.23+0.21}_{-0.24-0.22}$
\\
\hline
\end{tabular}
\caption{The values of $|V_{\rm ub}|$ in unit $10^{-3}$ for the WH-DA with $B\rhob=-0.2$, $0.0$ and $0.10$, respectively. The central values are obtained by setting all inputs to be their central values. The first (second) error is the squared average of the mentioned theoretical (experimental) uncertainties. } \label{Gamvub}
\end{table}

We put the predicted $|V_{\rm ub}|$ for the WH-DA in Table \ref{Gamvub}, where the first (second) error is the squared average of the mentioned theoretical (experimental) uncertainties. More specifically, the theoretical uncertainty comes from the choices of $b$-quark mass, the Borel window and the threshold parameter $s_0$; while, the experimental uncertainty comes from the errors of the measured lifetimes and decay ratios.

\begin{table}[htb]
\centering
\begin{tabular}{c  c | c  }
\hline
\multicolumn{2}{c|}{} & $\Vub$ \\ \hline
 \multicolumn{2}{c|}{$B\rhob=-0.2$} & ~~~~$2.91\pm0.19$~~~~  \\
\multicolumn{2}{c|}{ $B\rhob=0.0$}  & $3.11\pm0.19$  \\
\multicolumn{2}{c|}{ $B\rhob=+0.1$}  & $3.41\pm0.22$ \\
\hline
                                                    & LCSR~\cite{Ball04:Brho}  & $2.75\pm0.24$ \\
\raisebox {2.0ex}[0pt]{BABAR~\cite{BABAR:2010}}     & ISGW~\cite{ISGR}         & $2.83\pm0.24$ \\ \hline
                                                    & LCSR~\cite{Ball04:Brho}  & $2.85\pm0.40$ \\
\raisebox {2.0ex}[0pt]{BABAR~\cite{BABAR:2005}}     & ISGW~\cite{ISGR}         & $2.91\pm0.40$ \\
\hline
\end{tabular}
\caption{The weighted average of $\Vub$ in unit $10^{-3}$ from both the $B^+$-type and $B^0$-type and for the WH-DA with $B\rhob=-0.2$, $0.0$ and $0.10$, respectively. The estimations of the BABAR collaboration~\cite{BABAR:2010,BABAR:2005} are also presented as a comparison. } \label{Gammatota3}
\end{table}

Furthermore, we put the weighted average of the $B^+$-type and the $B^0$-type for $\Vub$ in Table \ref{Gammatota3}, in which the BABAR predictions are included as a comparison. The BABAR Collaboration predicts the $\Vub$ via the channel $B\to\rho\ell\nu$ based on two theoretical estimations on the $B\to\rho$ TFFs, i.e. the LCSR estimation of Ref.\cite{Ball04:Brho} and the ISGW estimation of Ref.\cite{ISGR}. Tables \ref{Gamvub} and \ref{Gammatota3} show that $|V_{\rm ub}|$ increases with the increment of $B^{\perp}_{2;\rho}$. So to compare with the BABAR predictions, the value of $B^{\perp}_{2;\rho}$ can not be too big.

\section{Summary}\label{sec:Summary}

The QCD LCSR has been used to deal with $B\to\rho$ TFFs, in which a chiral correlator has been suggested as the starting point. From the newly derived LCSRs, Eqs.(\ref{TFF_A1},\ref{TFF_A2},\ref{TFF_V}), we observe that the leading-twist LCDA $\phi_{2;\rho}^\bot$ provides the dominant contributions to the LCSRs for $B\to\rho$ TFFs. Thus, the uncertainties of the LCSRs from those uncertain high-twist LCDAs are highly suppressed. This makes $B\to \rho$ semi-leptonic decays be good places for testing the twist-2 $\phi_{2;\rho}^\bot$ models.

The WH-prescription provides a convenient way for constructing the LCWF/LCDA of the light mesons such as $\pi$, $K^{(*)}$ and $\rho$ mesons. More specifically, within the WH-prescription, the $\rho$-meson transverse momentum dependence is controlled by the BHL-prescription together with the Wiger-Melosh rotation effects, and its longitudinal distribution is dominantly controlled by a single parameter $B^{\perp}_{2;\rho}$. As a by-product, we can apply the $\rho$-meson LCWF into the $\rho$-meson involved processes under the pQCD factorization approach~\cite{pQCD}. Furthermore, we have presented the WH-model for $\rho$-meson LCDA in Eq.(\ref{DA:WH}). Varying $B_{2;\rho}^\bot$ from $-0.20$ to $+0.20$, the $\rho$-meson DA shall be varied from the single-peak behavior to the double-humped behavior, which covers most of the DA behaviors suggested in the literature. As examples, when taking $B_{2;\rho}^\bot \simeq -0.20$ and $0.10$, we obtain the same shape of CZ-DA and BB-DA.

The LCSRs for the $B\to\rho$ TFFs $A_{1}$, $A_{2}$ and $V$ have been discussed in detail under various LCDA models. Our present LCSRs agree with previous LCSRs derived via the conventional correlator as done by Ref.\cite{Ball04:Brho} but with less uncertainty. At present, the LCDAs' contributions at the $\delta^{1}$-order have been eliminated, thus we can draw more definite conclusions on the behavior of $\phi_{2;\rho}^\bot$. Table \ref{Endingpoint_1} shows the contributions from various LCDAs at the large recoil region. As required, it shows that the net twist-3 and twist-4 contributions are less than $10\%$ of the LCSRs, and the leading-twist LCDA $\phi_{2;\rho}^\bot$ do provide the dominant contributions.

After extrapolating $B\to\rho$ TFFs to their physical allowable region, we further make a comparison of them to those of lattice QCD calculations. The TFFs become smaller with the increment of $B_{2;\rho}^\bot$ and a larger $B_{2;\rho}^\bot$ (corresponding to a larger $a^\perp_2$) is not allowed by the lattice QCD predictions. For example, the AdS/QCD model, Eq.(\ref{phi_4}), with a much larger $B^\perp_{2;\rho}\sim 0.32$, could be excluded as indicated by Fig.(\ref{TFF:A1A2V}). If we have a more precise lattice QCD estimation, we can get a more strong constraint on the $\rho$-meson DA behavior.

We have applied the LCSRs for $B\to\rho$ TFFs to determine the total decay widths for $B\to\rho$ semileptonic decays and to determine the value of $|V_{\rm ub}|$. Table \ref{Gammatotal} shows the total decay width $\Gamma/|V_{\rm ub}^2|$ decreases and the $|V_{\rm ub}|$-free ratio $\Gamma^\|/\Gamma^\bot$ increases with the increment of $B^{\perp}_{2;\rho}$. Table \ref{Gamvub} shows that $|V_{\rm ub}|$ increases with the increment of $B^{\perp}_{2;\rho}$, thus, to compare with the BABAR prediction on $\Vub$, a larger $B^{\perp}_{2;\rho}$ is not allowable. For example, using the BABAR prediction based on the LCSR~\cite{BABAR:2010} as a criteria, we obtain $B^{\perp}_{2;\rho}\in[-0.2,0.10]$, which indicates that the $\rho$-meson LCDA prefers a single-peak behavior rather than a double-humped behavior. The $\rho$-meson LCDA shall be further constrained/tested by more data available in the near future, and we hope the definite behavior of $\rho$-meson LCDA can be concluded finally.

\hspace{2cm}

{\bf Acknowledgments}:  This work was supported in part by Natural Science Foundation of China under Grant No.11275280, by the Fundamental Research Funds for the Central Universities under Grant No.CQDXWL-2012-Z002, and by the Open Project Program of State Key Laboratory of Theoretical Physics, Institute of Theoretical Physics, Chinese Academy of Sciences, China under Grant NO.Y3KF311CJ1.

\appendix

\section{The $\rho$-meson high-twist LCDAs}
\label{sec:highertwist}

\begin{table}[htb]
\centering
\begin{tabular}{|c | c | c | c |}
\hline
$\mu$         &  $\zeta_3$        &  $\omega_3^A$  & $\omega_3^V$ \\\hline
$1{\rm GeV}$  &  $0.032\pm 0.010$ &  $-2.1\pm1.0$  & $3.8\pm1.8$  \\
$2.2{\rm GeV}$&  $0.018\pm 0.006$ &  $-1.7\pm0.9$  & $3.6\pm 1.7$ \\\hline
\end{tabular}
\caption{The parameters for the chiral-even 3-particle DAs.}
\label{tab:A}
\end{table}

\begin{table}[htb]
\centering
\begin{tabular}{|c | c | c | c | c | c |}
\hline
$\mu$  & $\omega_3^\bot$ & $\zeta_4^\bot$ & $\tilde{\zeta}_4^\bot$ & $\langle\langle Q^{(1)}\rangle\rangle$\\ \hline
1GeV   & $7.0\pm7.0$  & $0.10\pm 0.05$ & $-0.10\pm 0.05$ & $-0.15\pm 0.15$ \\
2.2GeV & $7.2\pm 7.2$ & $0.06\pm 0.03$ & $-0.06\pm 0.03$ & $-0.07\pm0.07$\\ \hline
\end{tabular}
\caption{The parameters for the chiral-odd 3-particle DAs.} \label{tab:B}
\end{table}

As has been shown in the body of the text, the twist-3 and twist-4 LCDAs at the $\delta^2$-order shall provide small contributions to the LCSRs. Thus, we directly adopt the expressions for those higher-twist DAs, i.e. $\phi_{3;\rho}^\|$, $\psi_{3;\rho}^\|$, $\phi_{4;\rho}^\bot$, $\psi_{4;\rho}^\bot$ and $\Phi_{3;\rho}^\bot$, suggested by Refs.\cite{P.Ball:1998-1999,Ball07:rhoWF}:
\begin{eqnarray}
\phi_{3;\rho}^\|(x)&=& 3\xi^2 + \frac{3}{2} a_2^\bot\xi^2 (5\xi^2-3)+\frac{15}{16}\zeta_{3}\omega_3^\bot(3\nn\\
&& -30\xi^2 +35\xi^4),\\
\psi_{3;\rho}^\|(x) & = & 6x\bar x \left[ 1 + \left( \frac{1}{4}a_2^\perp + \frac{5}{8}\zeta_3 \omega_3^\bot \right) (5\xi^2-1)\right], \\
\phi_{4;\rho}^\bot(x) && = 30 x^2 \bar x^2 \bigg\{ \frac{2}{5} \left( 1 +\frac{2}{7} a_2^\bot + \frac{10}{3}\zeta_4^\bot - \frac{20}{3} \wt{\zeta}_4^\bot \right) \nn\\
&& + \left( \frac{3}{35} a_2^\bot + \frac{1}{40}\zeta_3 \omega_3^\bot \right) C_2^{5/2}(\xi)\bigg\} -\bigg[\frac{18}{11}a_2^\bot \nn\\
&& - \frac{3}{2} \zeta_3 \omega_3^\bot + \frac{126}{55} \langle\!\langle Q^{(1)}\rangle\!\rangle + \frac{70}{11} \langle\!\langle Q^{(3)}\rangle\!\rangle \bigg]\nn\\
&&\times \bigg[ x\bar x (2+13 x\bar x) + 2x^3 (10-15x+6x^2)\nn\\
&&\times  \ln x  + 2\bar x^3 (10-15\bar x + 6\bar x^2) \ln \bar x\bigg],
\end{eqnarray}
\begin{eqnarray}
\psi_{4;\rho}^\bot(x) &=& 1 + \left\{\frac{3}{7} a_2^\perp -1 - 10 (\zeta_4^\bot + \wt{\zeta}_4^\bot ) \right\} C_2^{1/2}(\xi) \nn\\
&&+ \left[ -\frac{3}{7}\, a_2^\perp - \frac{15}{8} \zeta_3
\omega_3^\bot\right] C_4^{1/2}(\xi), \\
\Phi_{3;\rho}^\bot(\underline{\alpha}) &=& 540 \,\zeta_3\, \omega_3^\bot
(\alpha_1-\alpha_2) \alpha_1 \alpha_2 \alpha_3^2,
\end{eqnarray}
where the non-zero coefficients at two scales $1$GeV and $2.2$GeV are put in Tables \ref{tab:A} and \ref{tab:B}.

\end{document}